\documentclass[10pt]{iopart}

\usepackage{graphicx}
\usepackage{color}
\usepackage{bm}
\usepackage{cite}
\usepackage{multirow}
\usepackage{hyperref}
\usepackage{xspace}

\definecolor{green}{rgb}{0,0.6,0.1}

\newcommand{\xx}{$d_{x^2-y^2}$\xspace}

\usepackage{iopams}  
\begin{document}

\title{Superconductivity in infinite-layer nickelates}

\author{Yusuke Nomura$^1$, Ryotaro Arita$^{1,2}$}

\address{$^1$RIKEN Center for Emergent Matter Science, 2-1 Hirosawa, Wako, Saitama 351-0198, Japan \\
$^2$Department of Applied Physics, University of Tokyo, 7-3-1 Hongo Bunkyo-ku, Tokyo 113-8656, Japan}
\ead{yusuke.nomura@riken.jp}
\vspace{10pt}
\begin{indented}
\item[]June 2021
\end{indented}

\begin{abstract}
The recent discovery of the superconductivity in the doped infinite layer nickelates $R$NiO$_2$ ($R$=La, Pr, Nd) is of great interest since the nickelates are isostructural to doped (Ca,Sr)CuO$_2$ having superconducting transition temperature ($T_{\rm c}$) of about 110 K. Verifying the commonalities and differences between these oxides will certainly give a new insight into the mechanism of high $T_{\rm c}$ superconductivity in correlated electron systems. In this paper, we review experimental and theoretical works on this new superconductor and discuss the future perspectives for the ``nickel age'' of superconductivity.
\end{abstract}

%
%
%
%
\ioptwocol

\section{Introduction}
\label{sec_intro}

Unconventional superconductivity in strongly correlated electron systems is one of the most central issues in condensed matter physics. In particular, the mechanism of high-temperature superconductivity in cuprates~\cite{Bednorz_1986} has been the ``holy grail'' for more than 35 years, and many experimental and theoretical studies have been carried out. One promising strategy to unveil this long-standing puzzle is to compare cuprate superconductors with their variants and clarify which of the features of the cuprates plays a decisive role in superconductivity. However, the search for layered oxides with an electronic state similar to that of copper oxides is a highly non-trivial problem with a long history. In fact, a number of materials that are expected to be compatible with the keywords such as ``two-dimensional,'' ``square lattice,'' ``single orbital,'' and ``near half-filling'' have been investigated.

The cuprate parent compounds are a Mott insulator with $d^9$ filling in the copper $3d$ orbital (one hole per site in the hole picture). 
If there is particle-hole symmetry, a two-dimensional system with one $d$ electron per site will be promising.
Although titanium and vanadium oxides are candidates for a $d^1$ analog of cuprates, if we consider a similar crystal structure as cuprates, the electronic structure is not necessarily similar to that of cuprates: 
In the $d^1$ system in a typical octahedral crystal field, the electrons enter nearly degenerate $t_{2g}$ orbitals, and the ``single orbital'' condition, one of the important keywords for cuprate superconductivity, is not easily satisfied~\cite{Pickett_1989,Imai_2005,Matsuno_2005,Weng_2006}. 
Furthermore, the $d_{xy}$ level needs to be lower than that of the $d_{yz}$ and $d_{zx}$ orbitals in contrast to the typical octahedral crystal field, and it requires extremely high pressure to realize such a situation~\cite{Arita_2007}. 
It should also be noted that the $t_{2g}$ orbitals do not point towards their ligands, so that the electronic structure associated with them, as well as their exchange interactions, is very different from those of the $e_g$ orbitals.

In the cuprates, the onsite level of the $d_{x^2-y^2}$ orbital is higher than that of the $d_{3z^2-r^2}$ orbital. 
However, by lowering the position of the apical oxygen, the order can be reversed. 
Then a cuprate-like electronic state can be realized in the $d^7$ system with one electron in the $e_g$ shell. 
Nickel oxides provide a playground for designing such $d^7$ analogs.
However, the problem is how to control the position of the apical oxygen. 
Although such possibility has been explored considering artificial heterostructures, it seems difficult to achieve the ideal situation~\cite{Chaloupka_2008,Hansmann_2009,Han_2011}.

We can also consider a $d^5$ electron system where the orbital with the highest level in the $t_{2g}$ shell becomes half-filling.  
Iridium oxides are such an example, and there are experimental reports suggesting the possibility of superconductivity~\cite{Kim_2016,Yan_2015}. 
When the spin-orbit coupling is strong as in the $5d$ electron systems, the $t_{2g}$ bands split into $j_{\rm eff}=1/2$ and $j_{\rm eff}=3/2$ bands ($j_{\rm eff}$: effective total angular momentum).
The $j_{\rm eff}=1/2$ band becomes half-filling when the filling is $d^5$, and an electronic structure similar to that of the cuprates may be realized.
However, the energy splitting between the $j_{\rm eff}=1/2$ and $j_{\rm eff}=3/2$ bands is smaller than that between the $d_{x^2-y^2}$ and $d_{3z^2-r^2}$ orbitals in the cuprates, 
and it is not clear whether iridium oxides can be regarded as a  $j_{\rm eff}=1/2$ single-orbital system~\cite{MArtins_2011,Arita_2012,Zhang_2013}.

\begin{figure*}[tbp]
\vspace{0cm}
\begin{center}
\includegraphics[width=0.95\textwidth]{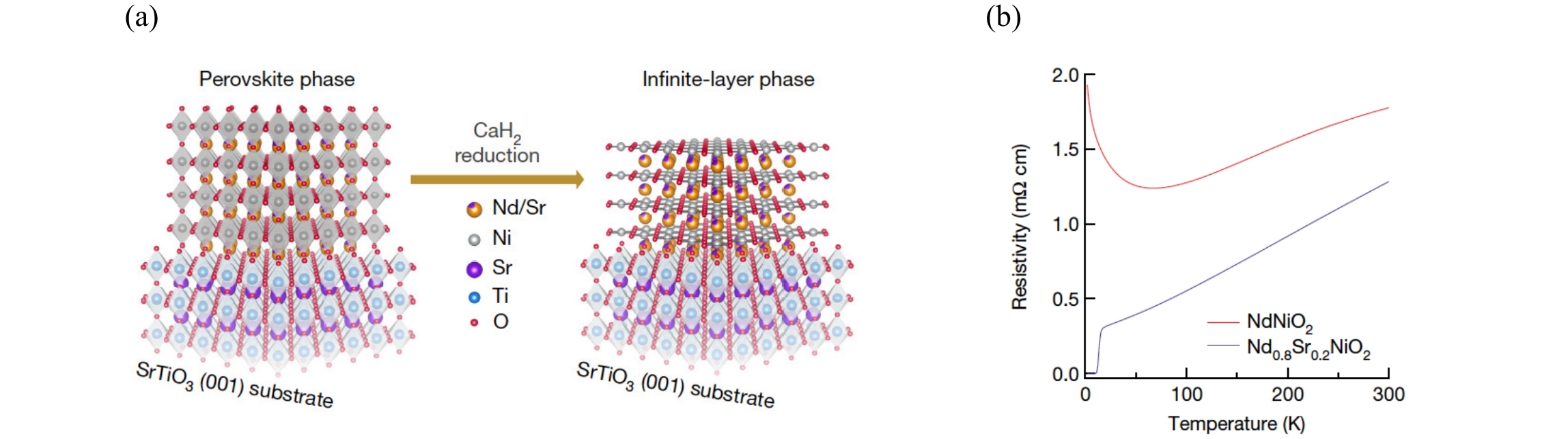}
\caption{
(a) Schematic figure for the crystal structures of Nd$_{0.8}$Sr$_{0.2}$NiO$_3$ (left) and Nd$_{0.8}$Sr$_{0.2}$NiO$_2$ (right) thin film samples on the SrTiO$_3$ substrate.
The doped infinite-layer nickelates are prepared by topotactic reduction. 
(b) Resistivity as a function of temperature for NdNiO$_2$ and Nd$_{0.8}$Sr$_{0.2}$NiO$_2$ thin film samples. 
Reproduced from Ref.~\cite{Li_2019}.
}
\label{Fig_discovery}
\end{center}
\end{figure*}

In this way, the hunt for analogs of cuprate superconductors has not been very successful so far. 
However, recently, there has been significant progress from infinite-layer nickelates. 
The infinite-layer phase of the nickelates is realized by topotactic reduction, whose first experimental report by Crespin {\it et al.}~\cite{Crespin_1983,Levitz_1983} (synthesis of LaNiO$_2$ powder) dates back to 1983, even before the discovery of the superconductivity in cuprates. 
Theoretically, in 1999, Anisimov {\it et al.}~\cite{Anisimov_1999} argued that infinite-layer nickelates could be a direct $d^9$ analog of the cuprates (to be precise, the $d$-orbital filling deviates from $d^9$, as described below), unlike the aforementioned attempts of $d^1$-, $d^7$-, and $d^5$-analogs. 
Finally, in 2019, Li {\it et al.} reported superconductivity with a transition temperature of 9--15 K in thin-film samples of doped infinite-layer nickelate Nd$_{0.8}$Sr$_{0.2}$NiO$_2$ (Fig.~\ref{Fig_discovery})~\cite{Li_2019}.  
The infinite-layer nickelate superconducting family has now extended to doped PrNiO$_2$~\cite{Osada_2020,Osada_2020_2} and LaNiO$_2$~\cite{Osada_2021,SW_Zeng_2022}. 
It should be noted that, although Sr$_2$RuO$_4$ also exhibits unconventional superconductivity and is isostructural to La$_2$CuO$_4$~\cite{Maeno_1994}, its electronic structure is very different from that of cuprates in that all the three $t_{2g}$ orbitals are involved with the formation of the Fermi surface.
Thus, it is of great interest to consider how the electronic structure of the new superconductor resembles and differs from that of the cuprates and how the similarity with cuprates can be improved.

In this review, in Sec.~\ref{sec_nickelate_basic}, we first compare the crystal structure and local electronic configuration of the infinite-layer nickelates and the cuprates.
 Then, we move on to the current status of the experiments in Sec.~\ref{sec_experiment}.
In Sec.~\ref{sec_ele_structure}, we discuss the electronic structure of the infinite-layer nickelates in more detail.
Sec.~\ref{sec_minimal_model} is devoted to the discussion on the essential degrees of freedom to describe the superconductivity. 
In Sec.~\ref{sec_new_materials}, we show several candidates for expanding the nickelate superconducting family. 
Finally, we give a summary and outlook in Sec.~\ref{sec_summary}.
For other recent reviews, see Refs.~\cite{Norman_2020,Pickett_2021,J_Zhang_2021,Botana_2021}.

\section{Crystal structure and crystalline electric field}
\label{sec_nickelate_basic}

\begin{figure}[tbp]
\vspace{0cm}
\begin{center}
\includegraphics[width=0.48\textwidth]{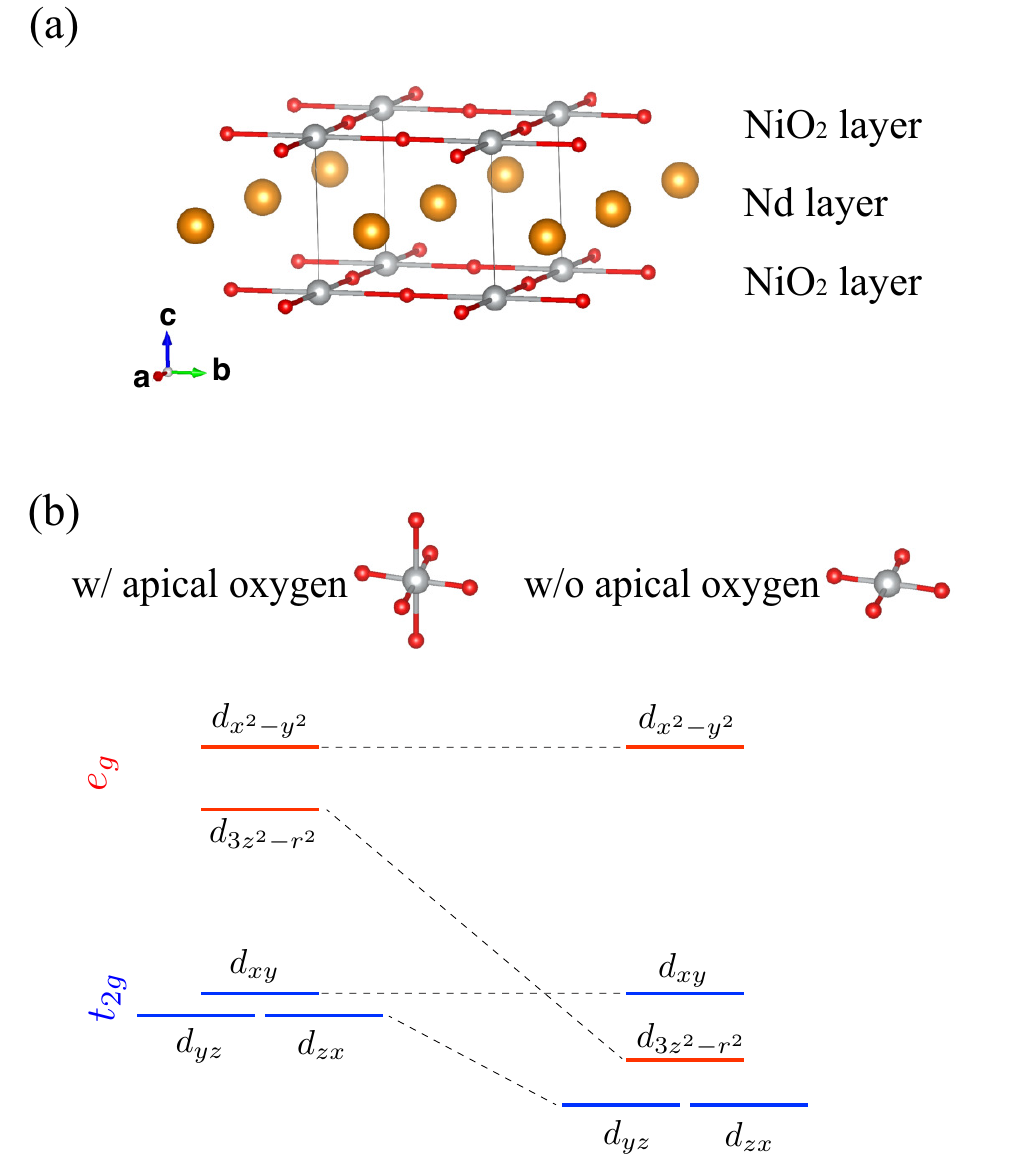}
\caption{
(a) The crystal structure of NdNiO$_2$, and (b) the corresponding square-planar crystal field (right). 
In (b), for comparison, the crystal field with apical oxygen realized in, e.g., La$_2$CuO$_4$ is also shown (left). 
We use VESTA~\cite{Momma_2011} to draw the crystal structure. 
}
\label{Fig_RNiO2}
\end{center}
\end{figure}

In this section, let us first look into the crystal structure and local electronic configuration of the infinite-layer nickelate $R$NiO$_2$ ($R\! =\!$ La, Pr, Nd) and compare with those of cuprates.
The crystal structure of $R$NiO$_2$ is shown in Fig.~\ref{Fig_RNiO2}(a).
It has a layered structure, called an infinite-layer structure, with alternating NiO$_2$ and $R$ layers. 
The structure is similar to that of the cuprates with a layered structure including CuO$_2$ planes. 
In fact, the corresponding infinite-layer cuprate (Sr$_{1-x}$Ca$_x$)$_{1-y}$CuO$_2$ also shows superconductivity with $T_{\rm c}$ of about 110 K~\cite{Azuma_1992}.

In the infinite-layer structure, the apical oxygens are absent, 
and the composition changes from the well-known nickel oxide $R$NiO$_3$.
A na\"ive valence estimate shows that neodymium is a $3+$ cation and oxygen is a $2-$ anion, giving nickel a $1+$ valence.
The corresponding occupation in the nickel $3d$ orbitals is $d^9$,  which is the same as that of the Cu$^{2+}$ cations in the cuprate parent compounds.
The square-planar crystal field [Fig. \ref{Fig_RNiO2}(b)] without the apical oxygens stabilizes the $d_{3z^2-r^2}$ orbital; its onsite level becomes comparable to those of the $d_{xy}$, $d_{yz}$, and $d_{zx}$ orbitals (called $t_{2g}$ orbitals in the octahedral environment). 
Therefore, the $d_{x^2-y^2}$ orbital is isolated from the other $3d$ orbitals in the energy diagram. 
Considering this crystal field together with the $d^9$ occupation, one expects that the half-filled $d_{x^2-y^2}$-orbital system is realized in $R$NiO$_2$.

As mentioned in Sec.~\ref{sec_intro}, ``two-dimensional,'' ``square lattice,'' ``single orbital,'' and ``near half-filling'' were the keywords to search for analogs of cuprates.
The infinite-layer nickelate $R$NiO$_2$ is a candidate that satisfies all of these.
The possible similarity between the nickelates and cuprates gives excitement in the field.
The ``nickel age'' of superconductivity has thus started~\cite{Norman_2020,Pickett_2021}.

\begin{figure*}[tbp]
\vspace{0cm}
\begin{center}
\includegraphics[width=1.00\textwidth]{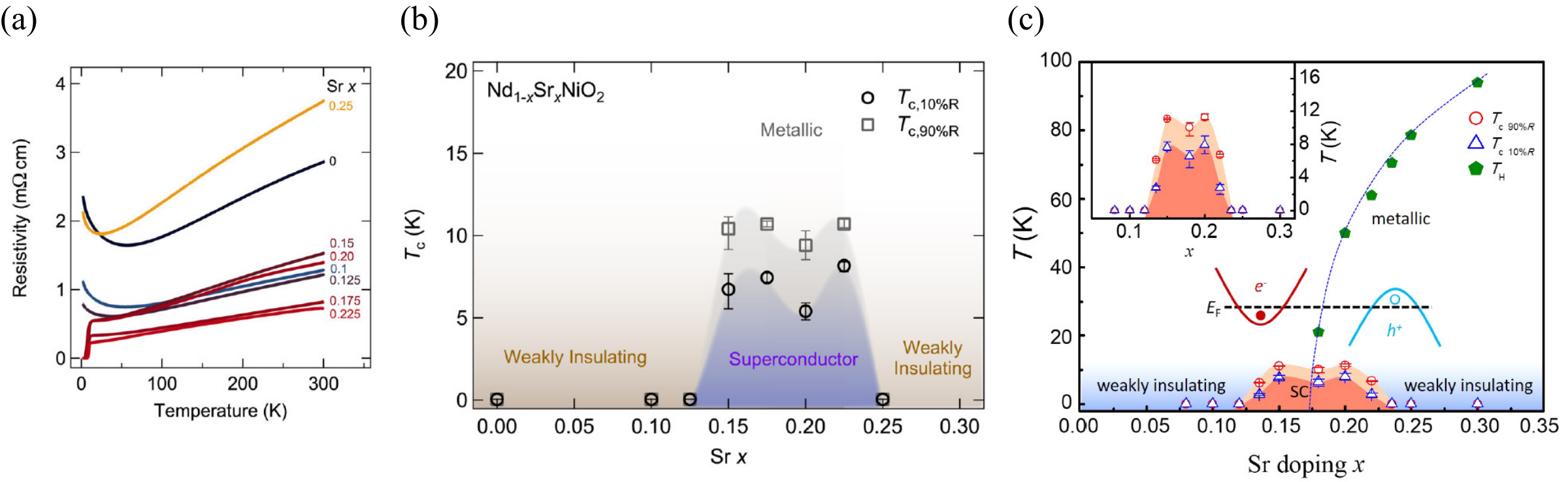}
\caption{
(a) Doping ($x$) dependence of the resistivity and (b,c) temperature-doping phase diagram of Nd$_{1-x}$Sr$_x$NiO$_2$ thin film samples on SrTiO$_3$ substrate.
The panels (a,b) and (c) are reproduced from Refs.~\cite{D_Li_2020} and \cite{Zeng_2020}, respectively. 
}
\label{Fig_phase_diagram}
\end{center}
\end{figure*}

\begin{figure*}[tbp]
\vspace{0cm}
\begin{center}
\includegraphics[width=1.00\textwidth]{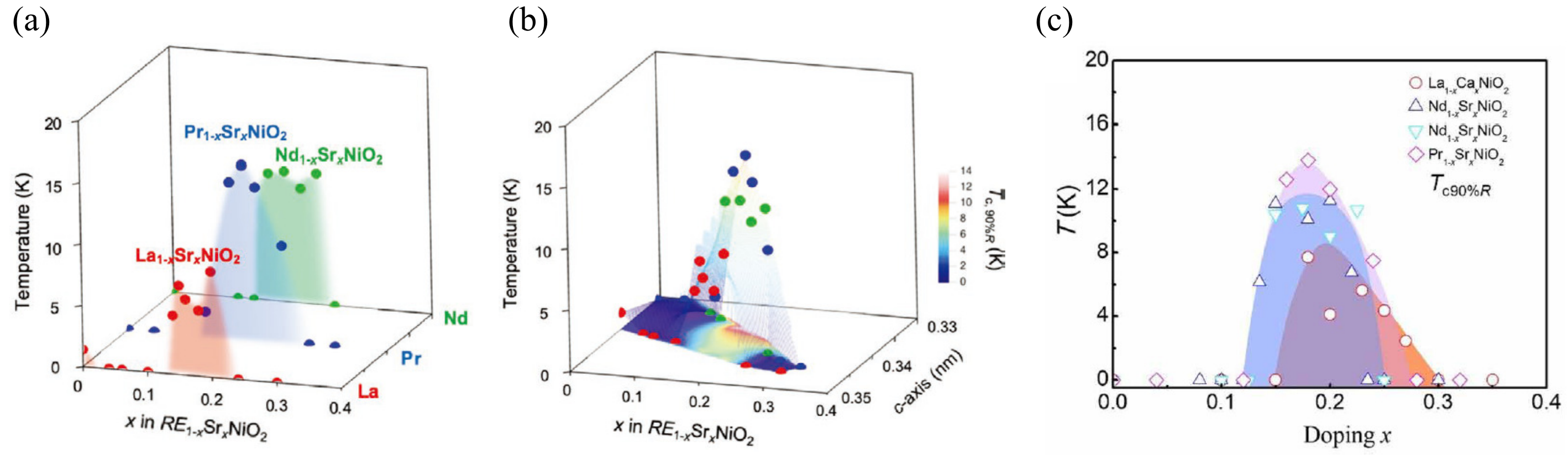}
\caption{
Comparison of the transition temperature among the superconducting infinite-layer nickelates [doped $R$NiO$_2$ ($R$ = La, Pr, Nd) thin film samples]. 
The panels (a,b) and (c) are reproduced from Refs.~\cite{Osada_2021} and \cite{SW_Zeng_2022}, respectively. 
}
\label{Fig_phase_diagram_comparison}
\end{center}
\end{figure*}

\section{Experimental studies}
\label{sec_experiment}

\subsection{Synthesis of superconducting materials}

In the first report~\cite{Li_2019}, D. Li {\it et al.} reported the superconductivity in the composition of Nd$_{0.8}$Sr$_{0.2}$NiO$_2$ ($T_{\rm c}=9$--15 K).
The superconductivity has also been observed in doped PrNiO$_2$ thin film, Pr$_{1-x}$Sr$_{x}$NiO$_2$, with a maximum $T_{\rm c}$ of 14 K~\cite{Osada_2020,Osada_2020_2}.
All the superconducting samples are thin film samples on SrTiO$_3$ substrates, and the details of the stable synthesis of film samples are discussed in Ref.~\cite{Lee_2020}.

It is worth noting that several other groups have now reproduced superconductivity in doped infinite-layer nickelate film samples~\cite{Zeng_2020,Q_Gu_2020,Q_Gao_2021,XR_Zhou_2021,Y_Li_2021}, although, in the early stage, there was a report of failure in reproducing the superconductivity in thin film samples~\cite{Zhou_2020}.
The phase diagram of Nd$_{1-x}$Sr$_{x}$NiO$_2$ thin film with different strontium concentration $x$ (Fig.~\ref{Fig_phase_diagram}) has been revealed independently by D. Li {\it et al.}~\cite{D_Li_2020} and S. W. Zeng {\it et al.}~\cite{Zeng_2020} (the groups of H. Y. Hwang and A. Ariando, respectively).
As the latest news, the same two groups have independently observed the superconductivity in doped LaNiO$_2$ thin films~\cite{Osada_2021,SW_Zeng_2022}.
This is remarkable because the initial work had not succeeded in realizing superconductivity in doped LaNiO$_2$~\cite{Li_2019} (see, e.g., Ref.~\cite{Liang_Si_2020} for the literature debating a possible reason for the absence of superconductivity).
Furthermore, Ref.~\cite{Osada_2021} also reported that there is a sign of superconductivity in undoped LaNiO$_2$ thin films [Fig.~\ref{Fig_phase_diagram_comparison}(a)]. 
This is interesting because NdNiO$_2$ and PrNiO$_2$ thin films do not show superconductivity in their parent compounds. 
In contrast, undoped LaNiO$_2$ film samples in Ref.~\cite{SW_Zeng_2022} are insulating from 300 K to 2K. 
It is desirable that the origin of the discrepancy between the two reports be clarified in the future.

As for bulk samples, to the best of our knowledge, there is no success in reproducing superconductivity. 
We refer to Refs.~\cite{Q_Li_2020,BX_Wang_2020,C_He_2021,Puphal_2021} for experimental effort on bulks (see also Ref.~\cite{Malyi_2022} for the discussion on the thermodynamical stability of the bulk infinite-layer nickelates).

\subsection{Phase diagram}

Figure~\ref{Fig_phase_diagram} shows the phase diagram of Nd$_{1-x}$Sr$_{x}$NiO$_2$ thin films on the SrTiO$_3$ substrate. 
First, the resistivity of the parent material does not show a clear insulating behavior.
The term ``weakly insulating'' refers to the slight upturn of the resistivity at low temperatures.
Superconductivity is observed in the region where the strontium doping concentration is $0.125 \lesssim x \lesssim 0.25$, and a dome-shaped $T_{\rm c}$ is observed.
In the overdoped region, where the superconductivity disappears, the resistivity of Nd$_{1-x}$Sr$_{x}$NiO$_2$ increases at low temperatures, in contrast to the cuprates, which show metallic behavior. 
The increase of the resistivity may be due to, e.g., disorder, electron correlation, and secondary order parameters. 
The origin of this behavior is discussed in, e.g., Ref.~\cite{Hsu_2021} (see Sec.~\ref{sec_experiment_indivisual}).

What is striking is the difference in the behavior of the parent materials between the nickelate and the cuprates. 
The cuprate parent compounds are an antiferromagnetic Mott insulator, whereas NdNiO$_2$ is not an insulator. 
Whereas the presence of the magnetic order in the thin film samples is not clear, the bulk powder NdNiO$_2$ samples show no long-range antiferromagnetic ordering down to low temperatures (1.7 K)~\cite{Hayward_2003}. 
We will discuss magnetism further in Sec.~\ref{sec_magnetism}.

Figure~\ref{Fig_phase_diagram_comparison} shows the comparison of $T_{\rm c}$ among the superconducting infinite-layer nickelates~\cite{Osada_2021,SW_Zeng_2022}.
One may be able to draw a unified phase diagram like in Fig.~\ref{Fig_phase_diagram_comparison}(b).
It is an important future task to clarify the similarities and differences in more detail among superconducting family members.

\subsection{Experiments to understand electronic structure, superconductivity, and magnetism}
\label{sec_experiment_indivisual}

Other experimental facts are summarized as follows.

\begin{figure}[tbp]
\vspace{0cm}
\begin{center}
\includegraphics[width=0.48\textwidth]{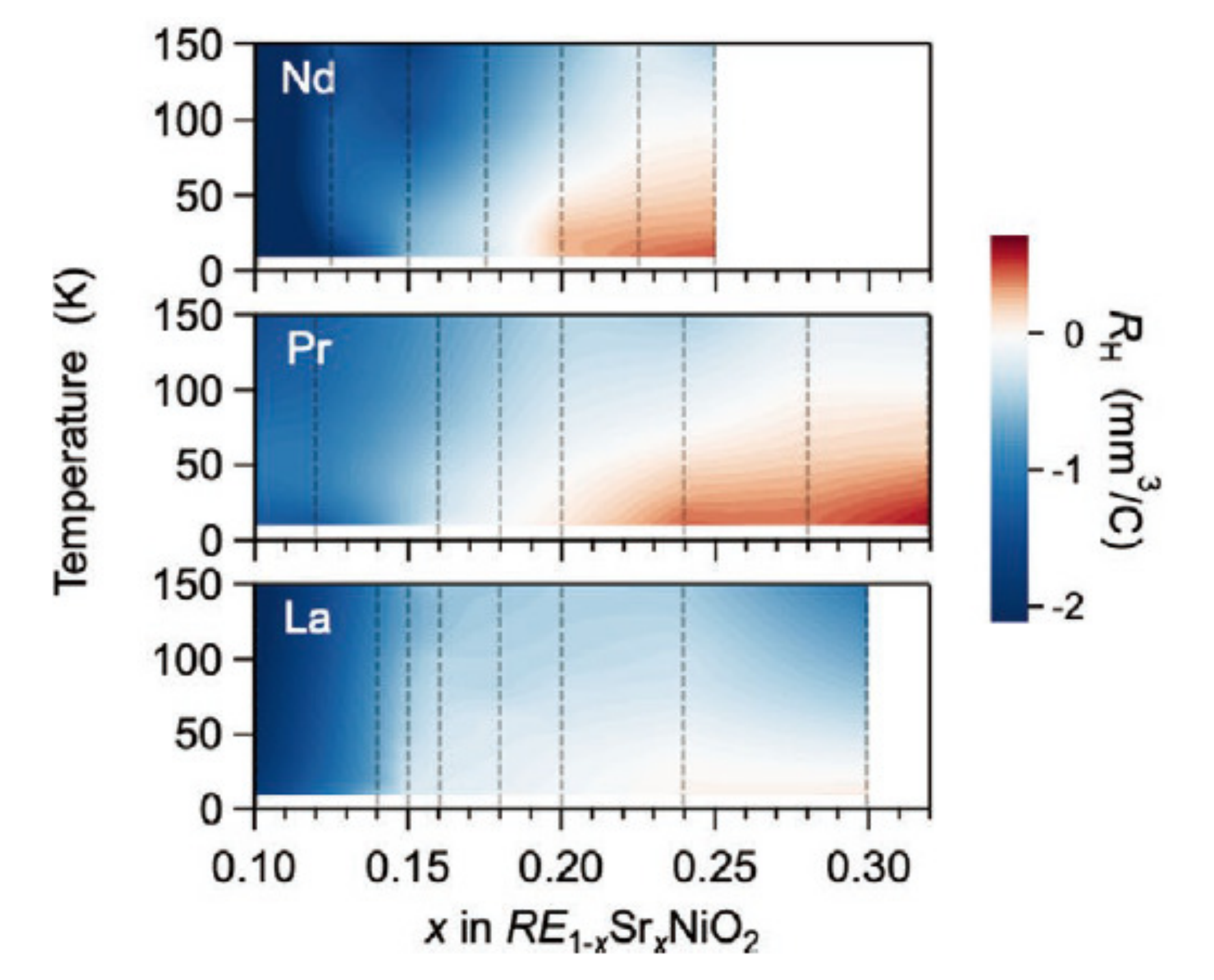}
\caption{
Contour maps of the Hall coefficients of doped NdNiO$_2$ PrNiO$_2$, and LaNiO$_2$ thin films.
Reproduced from Ref.~\cite{Osada_2021}.
}
\label{Fig_Hall}
\end{center}
\end{figure}

\begin{itemize}
    \item Hall coefficient measurements of thin film samples suggest a coexistence of electron and hole carriers in Nd$_{1-x}$Sr$_{x}$NiO$_2$, and the sign of the Hall coefficient changes depending on the strontium concentration $x$ and the temperature~\cite{D_Li_2020,Zeng_2020}. 
    A qualitatively similar sign change is also seen in doped PrNiO$_2$ and LaNiO$_2$ thin films~\cite{Osada_2020_2,Osada_2021,SW_Zeng_2022} (Fig.~\ref{Fig_Hall}).

    \item Experiments using the $X$-ray techniques and electron energy loss spectroscopy (EELS) reveal that the electronic states of the rare-earth layer show up around the Fermi level [Fig.~\ref{Fig_Xray-EELS}(a)] and that NdNiO$_2$ is more close to the Mott-Hubbard regime in Zaanen-Sawatzky-Allen classification~\cite{Zaanen_1985} compared to cuprates~\cite{Hepting_2020,Fu_arXiv,Goodge_2021,Z_Chen_arXiv}.
    It suggests that the holes are mainly doped into nickel $3d$ orbitals (a signal for the doped holes in the oxygen $2p$ orbitals is also observed in the O $K$-edge EELS measurement, but the intensity is much smaller than that of the cuprates~\cite{Goodge_2021} [Fig.~\ref{Fig_Xray-EELS}(b)]). 

\begin{figure}[tbp]
\vspace{0cm}
\begin{center}
\includegraphics[width=0.5\textwidth]{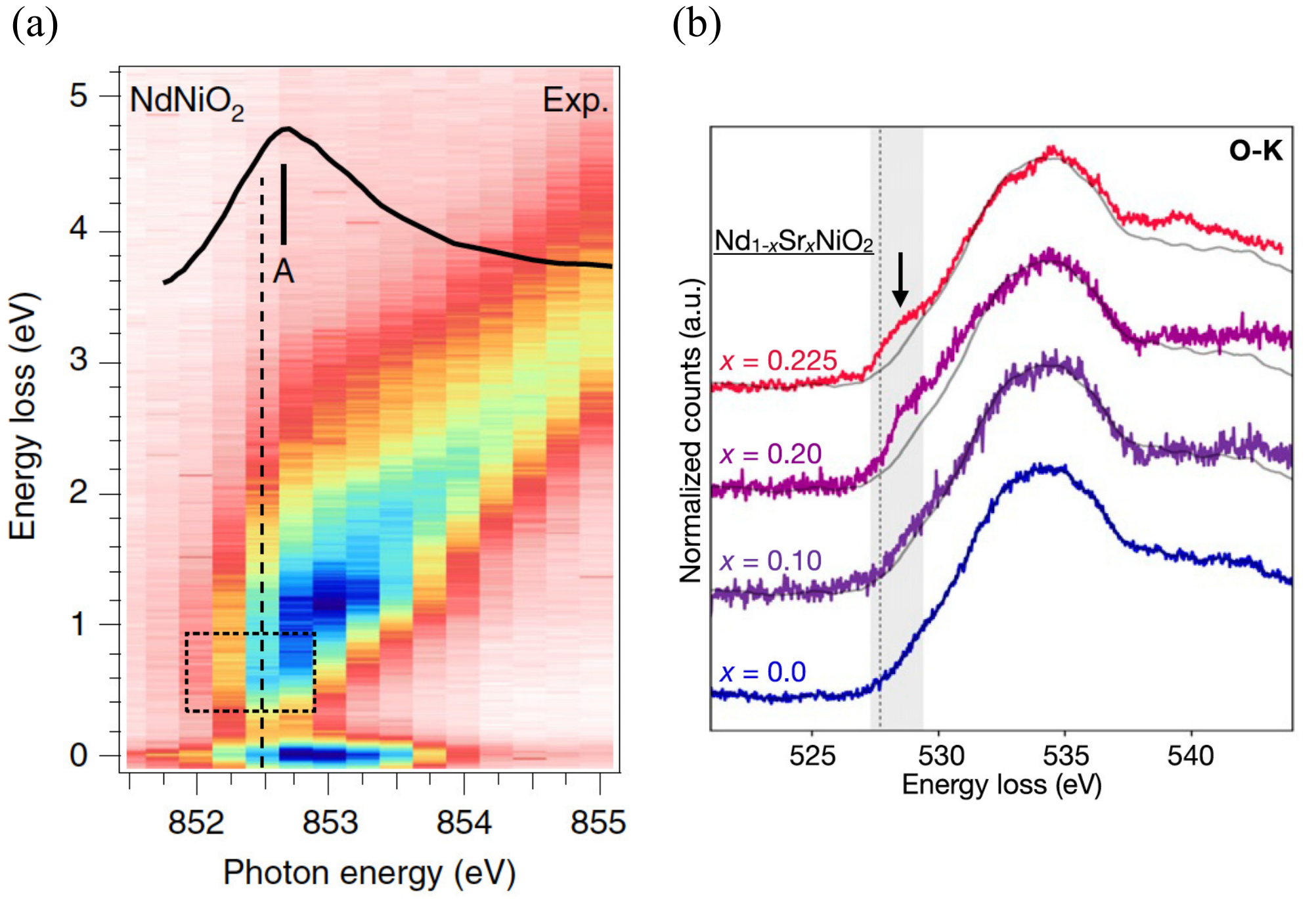}
\caption{
(a) Ni L$_3$-edge XAS (X-ray absorption spectroscopy) and RIXS (resonant inelastic X-ray scattering) of NdNiO$_2$ thin films. 
The XAS result is superimposed in the RIXS intensity map. 
The feature highlighted by the dashed box is interpreted as a signal for the Ni-Nd hybridization in Ref.~\cite{Hepting_2020}. 
(b) O K-edge EELS of Nd$_{1-x}$Sr$_{x}$NiO$_2$ thin films.
A feature at around 528 eV in doped samples is attributed to $d^9 \underline{L}$ states in Ref.~\cite{Goodge_2021}. 
Reproduced from (a) Ref.~\cite{Hepting_2020} and (b) Ref.~\cite{Goodge_2021}. 
}
\label{Fig_Xray-EELS}
\end{center}
\end{figure}

    \item Motivated by the discovery of the superconductivity, the magnetism of the bulk infinite-layer nickelates is reinvestigated using NdNiO$_2$~\cite{Fu_arXiv,H_Lin_2021}, Nd$_{0.85}$Sr$_{0.15}$NiO$_2$~\cite{Y_Cui_2021}, LaNiO$_2$~\cite{D_Zhao_2021,H_Lin_2021}, and PrNiO$_2$~\cite{H_Lin_2021} samples. Recently, a RIXS measurement has been performed on NdNiO$_2$ thin film samples~\cite{Lu_2021}. A quasi-two-dimensional magnetic excitation with a bandwidth of about 200 meV with large damping is observed (Fig.~\ref{Fig_spin_excitation}). It suggests a nonnegligible magnetic exchange interaction of about 64 meV on the NiO$_2$ layer~\cite{Lu_2021}, while a Raman experiment using bulk samples gave a smaller value of 25 meV~\cite{Fu_arXiv}.
    The origin of the lack of a clear long-range magnetic order is also an open question. See Sec.~\ref{sec_magnetism} for more detailed discussion. 

\begin{figure}[tbp]
\vspace{0cm}
\begin{center}
\includegraphics[width=0.48\textwidth]{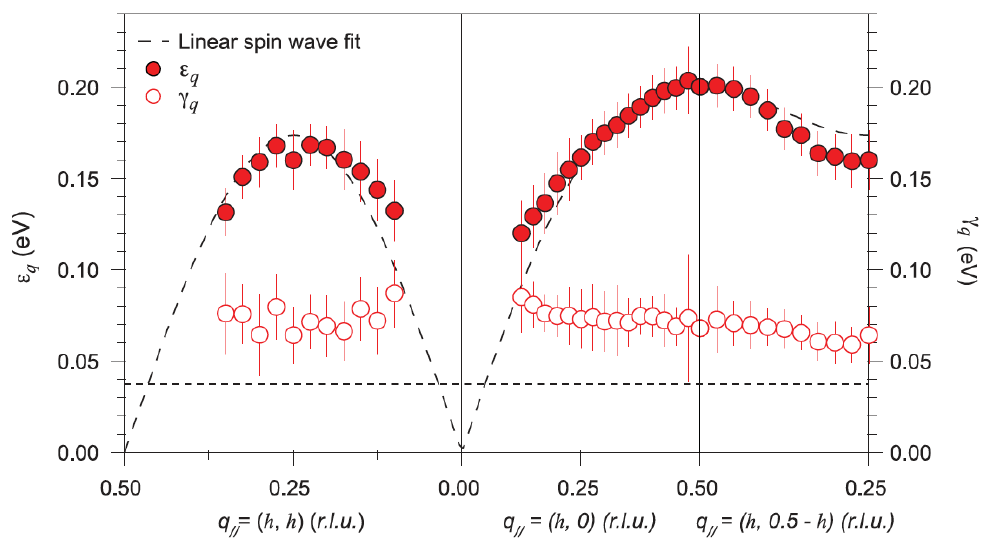}
\caption{
Dispersion of magnetic excitation in NdNiO$_2$ thin films inferred from the RIXS. The filled and open red circles indicate the energy of magnetic mode and damping factor, respectively. 
The dashed curve is the linear spin wave dispersion assuming the two-dimensional Heisenberg model with the nearest-neighbor coupling $J_1 = 63.6 \pm 3.3 $ meV and the next-nearest-neighbor coupling $J_2=-10.3 \pm 2.3$ meV. Reproduced from Ref.~\cite{Lu_2021}.}
\label{Fig_spin_excitation}
\end{center}
\end{figure}

    \item Scanning tunneling microscope/spectroscopy (STM/STS) experiments on Nd$_{1-x}$Sr$_{x}$NiO$_2$ film samples observe, depending on the position of the inhomogeneous surface, $V$-shape or full-gap-type single-particle tunneling spectra (Fig.~\ref{Fig_STS}). In some cases, mixed spectra of the two components are observed~\cite{Q_Gu_2020}. The $V$-shape spectrum is compatible with the $d$-wave gap, whereas the full-gap-type spectrum suggests the $s$-wave symmetry. 
    Ref.~\cite{Q_Gu_2020} interprets the coexistence of the different tunneling spectra by the different gap symmetries on different Fermi surfaces, while other explanations may also be plausible. The effects of the multi-orbital gap structure, the inhomogeneity, and surface reconstruction remain to be elucidated~\cite{Adhikary_2020,Z_Wang_2020,Kitamine_2020,X_Wu_arXiv}.

    \item The upper critical field of the doped NdNiO$_2$ thin film samples is investigated by two independent groups~\cite{BY_Wang_2021,Y_Xiang_2021}. Compared to the cuprates, the upper critical field shows a smaller spatial anisotropy. The analyses of the critical field suggest that the paramagnetic effect is dominant over the orbital effect (Pauli limit).

\begin{figure}[tbp]
\vspace{0cm}
\begin{center}
\includegraphics[width=0.5\textwidth]{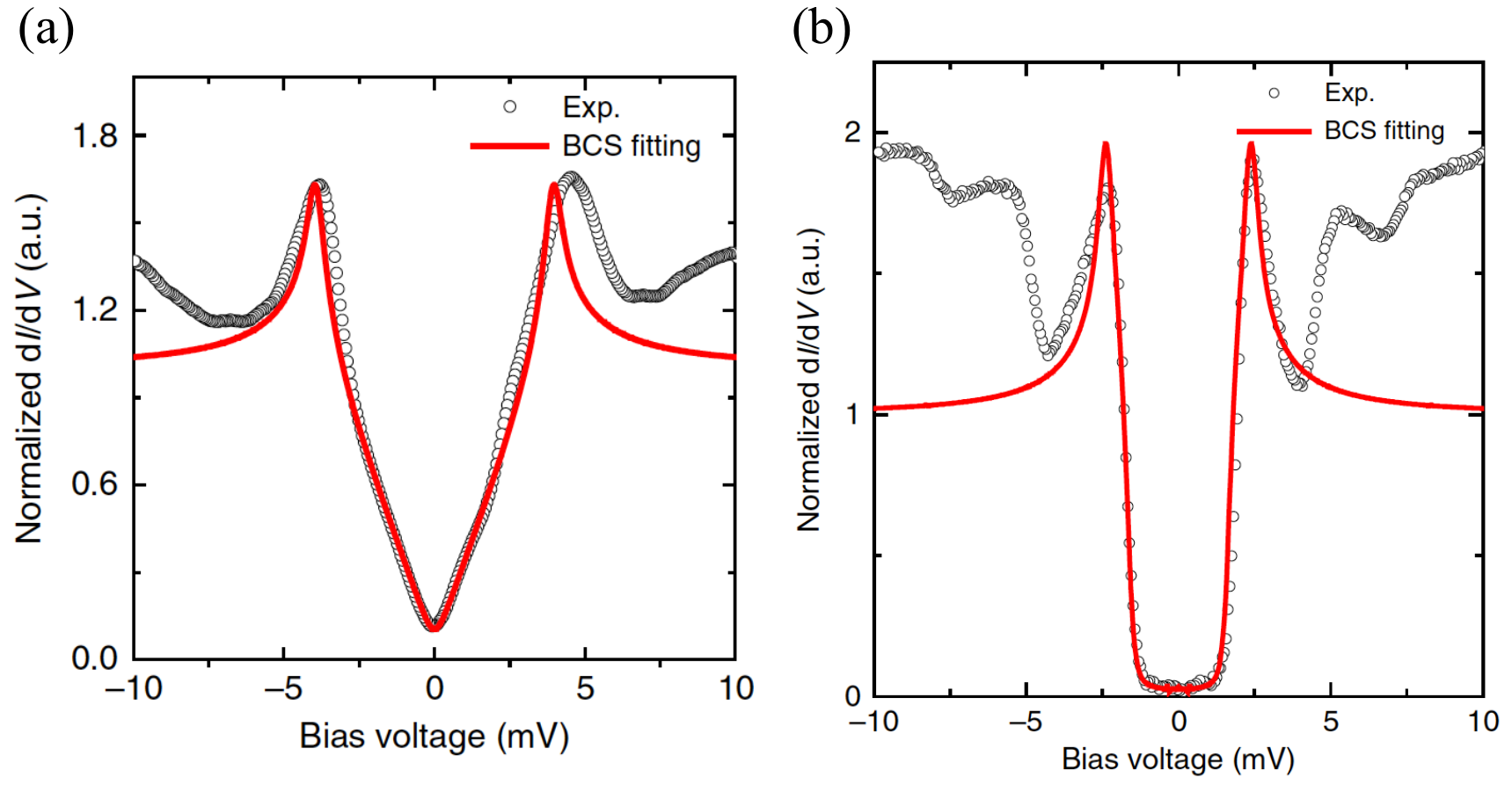}
\caption{
(a) A typical $V$-shape tunneling spectrum (black circles)
and (b) a full-gap-type tunneling spectrum (black circles)
measured at the surfaces of Nd$_{1-x}$Sr$_x$NiO$_2$ thin films.
The spectrum depends on the position of the STM tip~\cite{Q_Gu_2020}. 
Reproduced from Ref.~\cite{Q_Gu_2020}.
}
\label{Fig_STS}
\end{center}
\end{figure}

    \item When the system is in the Mott-Hubbard regime, the doped configuration becomes mainly $d^8$, rather than $d^9 \underline{L}$ ($\underline{L}$ denotes a hole in ligand oxygen) in the case of the charge-transfer regime. Ref.~\cite{Rossi_2021} has investigated the multiplet structure of the $d^8$ configuration of doped NdNiO$_2$ film samples using the XAS and RIXS, and concluded that the orbital-polarized spin-singlet state, where the doped holes reside in the nickel 3\xx orbital, gives a dominant contribution. 
    See Sec.~\ref{sec_other_3d} for the detailed discussion on the $d^8$ multiplet structure.
    
    \item Ref.~\cite{SW_Zeng_2022_2} has reported a diamagnetic response in superconducting Nd$_{1-x}$Sr$_{x}$NiO$_2$ film samples. 
    The film thickness dependence of the physical observables is also explored. 
    The superconductivity is observed for different thickness (from 4.6 nm to 15.2 nm) samples, and thicker films tend to show a higher $T_{\rm c}$.
    The change in the Hall coefficient and the XAS spectra depending on the thickness implies the modulation of the electronic structure due to the strain and interface effects. 
    
    \item Normal state resistivity of Nd$_{1-x}$Sr$_{x}$NiO$_2$ film samples is studied in Ref.~\cite{Hsu_2021} by suppressing the superconductivity with out-of-plane magnetic fields. The upturn of the resistivity at low temperatures observed around $x=0$ outside the superconducting dome persists in the field-induced normal state up to $x \approx 0.225$. At $x \approx 0.225$, a metallic transport is observed, but, above $x \approx 0.225$, the resistivity upturn shows up again. The systematic doping evolution of the transport property implies the intrinsic correlation effect and possible secondary order parameter, for which further investigations are required. 
\end{itemize}
The consistency between these experimental facts and theory will be discussed in the following sections. 

\begin{figure*}[tbp]
\vspace{0cm}
\begin{center}
\includegraphics[width=0.95\textwidth]{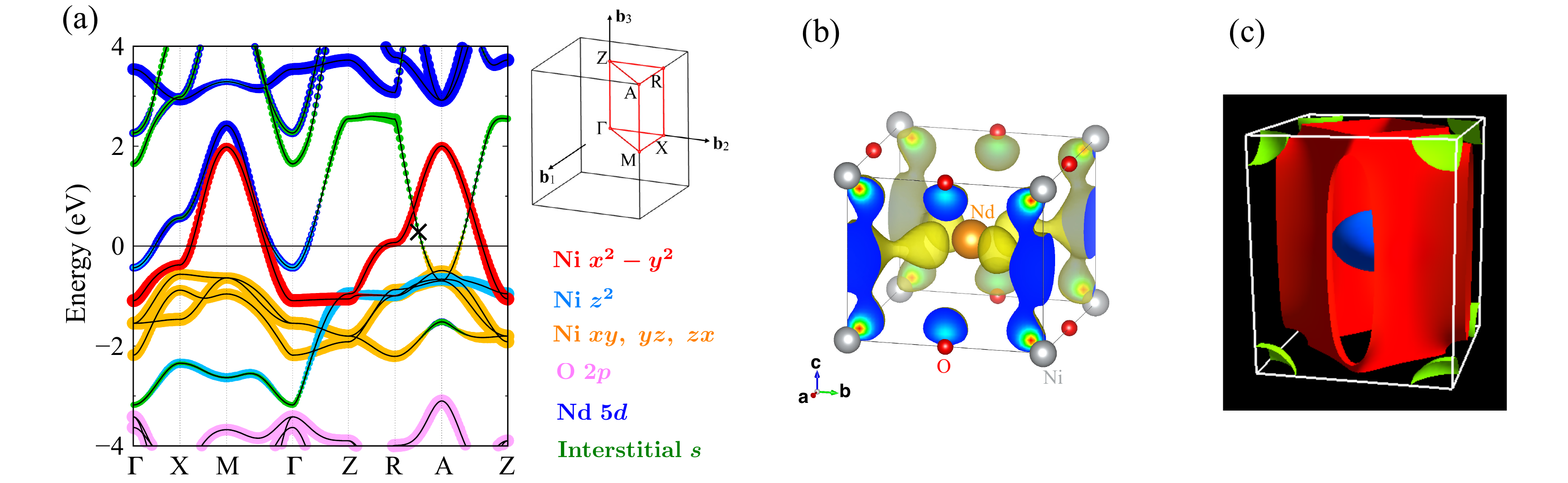}
\caption{
Electronic structure of bulk NdNiO$_2$ calculated using DFT with the GGA exchange-correlation functional. 
The lattice constants are taken from Ref.~\cite{Hayward_2003}.
(a) Band structure of NdNiO$_2$ colored by the weight of each orbital (fat band).
The neodymium $4f$ orbitals, which would form localized spins, are treated as ``frozen core''. 
(b) Contour plot of the real-space electron density of the Bloch state marked with the $\times$ symbol between the $R$ and $A$ points near the Fermi level in the panel (a). 
Orange, gray, and red spheres indicate the neodymium, nickel, and oxygen atoms, respectively.  
Yellow surfaces show an isosurface of the electron density.
It has large weights around the missing apical oxygen site and neodymium site. 
Reproduced from Ref.~\cite{Nomura_2019}.
(c) Fermi surface of NdNiO$_2$. 
For the labels of high-symmetry $k$ points in the Brillouin zone, see the panel (a).
The nickel $3d_{x^2-y^2}$ orbital forms a large quasi-two-dimensional Fermi surface (red). 
The Fermi pocket around the $\Gamma$ point (blue) has mainly neodymium $5d_{3z^2-r^2}$ character. 
The Fermi pocket near the $A$ point (green) is derived from the bonding orbital between the interstitial $s$ orbital and the neodymium $5d_{xy}$ orbital [see (b)].
The image was drawn using FermiSurfer\cite{Kawamura_2019}.
}
\label{Fig_band_NdNiO2}
\end{center}
\end{figure*}

\section{Electronic structure of infinite-layer nickelates}
\label{sec_ele_structure}

\subsection{Insight from DFT calculations}

In Sec.~\ref{sec_nickelate_basic}, we have discussed that the parent material, $R$NiO$_2$ ($R$=La, Pr, Nd), might be a single-orbital strongly-correlated system on a two-dimensional square lattice with half-filled $3d_{x^2-y^2}$ orbital, based on simple valence and crystal-field analyses.
However, as we discuss in the following, this picture is not entirely true~\cite{Lee_2004,Botana_2020,Sakakibara_2020,Wu_2020,Nomura_2019}.
Experimental facts reviewed in Sec.~\ref{sec_experiment} also suggest that there should be a deviation from the simple picture. 
In this section, we discuss the electronic structure of the infinite-layer nickelates through first-principles calculations.

Here, we restrict ourselves to bulk properties. 
The film thickness is on the order of 10 nm~\cite{Lee_2020}, so that there are several tens of NiO$_2$ layers in the sample. 
Furthermore, the superconductivity is robustly observed for samples with different thicknesses (from 4.6 nm to 15.2 nm)~\cite{SW_Zeng_2022_2}. 
However, since the infinite-layer structure consists of charged layers, the reconstruction of the lattice and electronic structures is naturally expected at the interfaces and surfaces. 
Such effect is studied theoretically in Refs.~\cite{Bernardini_2020c,Geisler_2020,Geisler_2021,He_2020,Y_Zhang_2020,X_Wu_arXiv} (see also Ref.~\cite{Botana_2021}). 
Experimental consideration has also started~\cite{SW_Zeng_2022_2} (see Sec~\ref{sec_experiment_indivisual}).

\begin{figure*}[tbp]
\vspace{0cm}
\begin{center}
\includegraphics[width=0.95\textwidth]{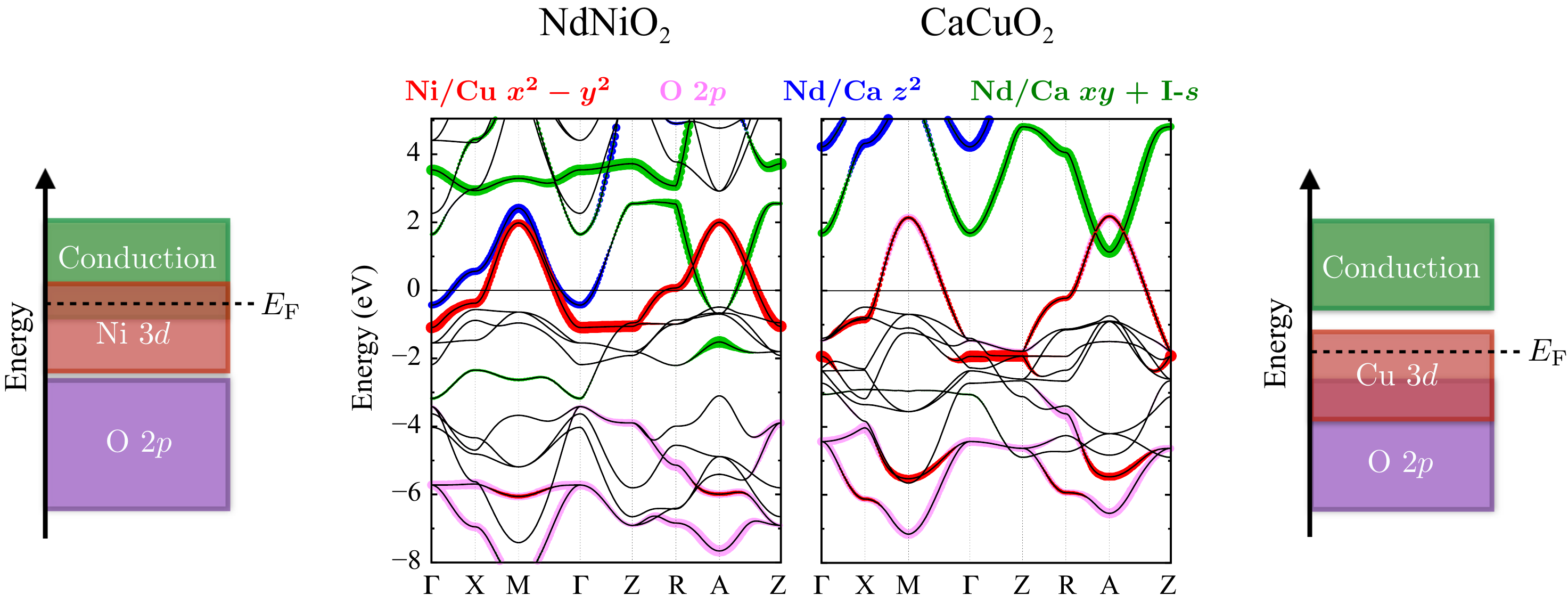}
\caption{
Comparison of the DFT-GGA band structures between NdNiO$_2$ and CaCuO$_2$.
Weights of nickel/copper $3d_{x^2-y^2}$ orbital, oxygen $2p$ orbitals hybridized with $3d_{x^2-y^2}$ orbitals in NiO$_2$/CuO$_2$ planes, neodymium/calcium $d_{3z^2-r^2}$ orbitals, and sum of neodymium/calcium $d_{xy}$ and interstitial $s$ (I-$s$ orbital, I stands for interstitial) contributions are shown as fat bands.
Note that the fat bands for the oxygen $2p$ orbitals here are limited to those hybridized with the $3d_{x^2-y^2}$ orbitals ($p_x$ orbital of the oxygen in the $x$-direction and $p_y$ orbital of oxygen in the $y$-direction), whereas, in Fig.~\ref{Fig_band_NdNiO2}, the fat bands were drawn for all the oxygen $2p$ orbitals. 
The lattice parameters of CaCuO$_2$ are taken from Ref.~\cite{Siegrist_1988}.
The figure is created inspired by the comparison between LaNiO$_2$ and CaCuO$_2$ shown in Refs.~\cite{Lee_2004,Botana_2020}. 
}
\label{Fig_NdNiO2_CaCuO2}
\end{center}
\end{figure*}

Figure~\ref{Fig_band_NdNiO2}(a) shows the band structure of the parent material NdNiO$_2$ (the first infinite-layer nickelate superconductor) calculated based on the density functional theory (DFT) with the generalized gradient approximation (GGA).
In general, the band structure of strongly correlated materials deviates (at least quantitatively) from that of the DFT calculation due to electron correlation effects (mass renormalization, Mott-gap opening, and so on), especially near the Fermi level~\cite{Held_2007}. However, to see the global energy structure, the DFT-GGA calculations give a good starting point.
See Sec.~\ref{sec_beyond_DFT} for the discussion of correlation effects beyond the DFT.

The analysis of the band character near the Fermi level (Fig.~\ref{Fig_band_NdNiO2}) shows that the nickel $3d_{x^2-y^2}$ orbital creates a large Fermi surface.
This is in agreement with the prediction of Sec.~\ref{sec_nickelate_basic}, and is a common feature with the cuprates. 
Also, consistently with the crystal-field analysis in Sec.~\ref{sec_nickelate_basic}, the $3d_{x^2-y^2}$ orbital has the highest onsite level among the nickel $3d$ orbitals, and the $3d_{x^2-y^2}$ band is isolated from the other $3d$ bands in the energy space.

However, there is a discrepancy from the na\"ive expectation shown in Sec.~\ref{sec_nickelate_basic}.  
Besides the large Fermi surface of the nickel $3d_{x^2-y^2}$ orbital, additional Fermi pockets exist near the $\Gamma$(0,0,0) and A($\pi/a,\pi/a, \pi/c$) points [Figs.~\ref{Fig_band_NdNiO2}(a) and \ref{Fig_band_NdNiO2}(c)].
The Fermi pocket around the $\Gamma$ point is formed by the rare-earth $5d_{3z^2-r^2}$ orbital with nickel $3d_{3z^2-r^2}$ orbital being hybridized~\cite{Lee_2004,Botana_2020}. 
This fact may make nickel $3d_{3z^2-r^2}$ orbital active at the parent compound~\cite{Lee_2004}. 

The origin of the Fermi pocket around the $A$ point is referred to as the rare-earth $5d_{xy}$ orbital~\cite{Sakakibara_2020,Wu_2020} or the interstitial $s$ orbital at the apical site~\cite{Y_Gu_2020}.
In reality, the bonding orbital is formed between these two orbitals~\cite{Nomura_2019} [Fig.~\ref{Fig_band_NdNiO2}(b)]: 
At the $A$ point, there is a large bonding-antibonding energy splitting of more than 10 eV~\cite{Hirayama_2020}, and the bonding part forms the additional Fermi pocket. 
This bonding orbital can be described either by $5d_{xy}$-like Wannier orbital centered at the rare-earth site or the $s$-like orbital centered at the interstitial apical site; therefore, both pictures can be applied~\cite{Nomura_2019}.

In most materials, interstitial-orbital bands appear far away from the Fermi level on the unoccupied side.  However, in NdNiO$_2$, the interstitial orbital at the apical site is stabilized because there is a space to gain the kinetic energy thanks to the absence of apical oxygen, and negatively-charged electrons feel an attraction from the surrounding nickel and neodymium cations~\footnote{The band structure is similar to that of electrides, in which interstitial bands are occupied and the electrons themselves become negative ions.}. 
The neodymium $5d$ and interstitial-$s$ orbitals are both located in the neodymium layer, and will henceforth be referred to collectively as Nd-layer orbitals (or more generally, rare-earth-layer orbitals).

Because the rare-earth-layer orbitals form the additional Fermi pockets, the nickel $3d_{x^2-y^2}$ orbital deviates from the half-filling even in the parent compound (occupation of the nickel $3d$ orbitals is not $d^9$)~\cite{Lee_2004}.
This ``self-doping'' effect is also seen in other infinite-layer compounds $R$NiO$_2$ (see, e.g., Refs.~\cite{Been_2021,Kapeghian_2020} for systematic investigation on the rare-earth element dependence). 
The self-doping marks one of the major differences from the cuprates, in which only the $3d_{x^2-y^2}$ band crosses the Fermi level (except for some materials)~\cite{Lee_2004}.
The DFT-level estimates suggest that the nickel $3d_{x^2-y^2}$ orbital is about 10 \% hole doped at the parent compound~\cite{Botana_2020,Sakakibara_2020,Nomura_2019}.

\begin{figure}[tbp]
\vspace{0cm}
\begin{center}
\includegraphics[width=0.48\textwidth]{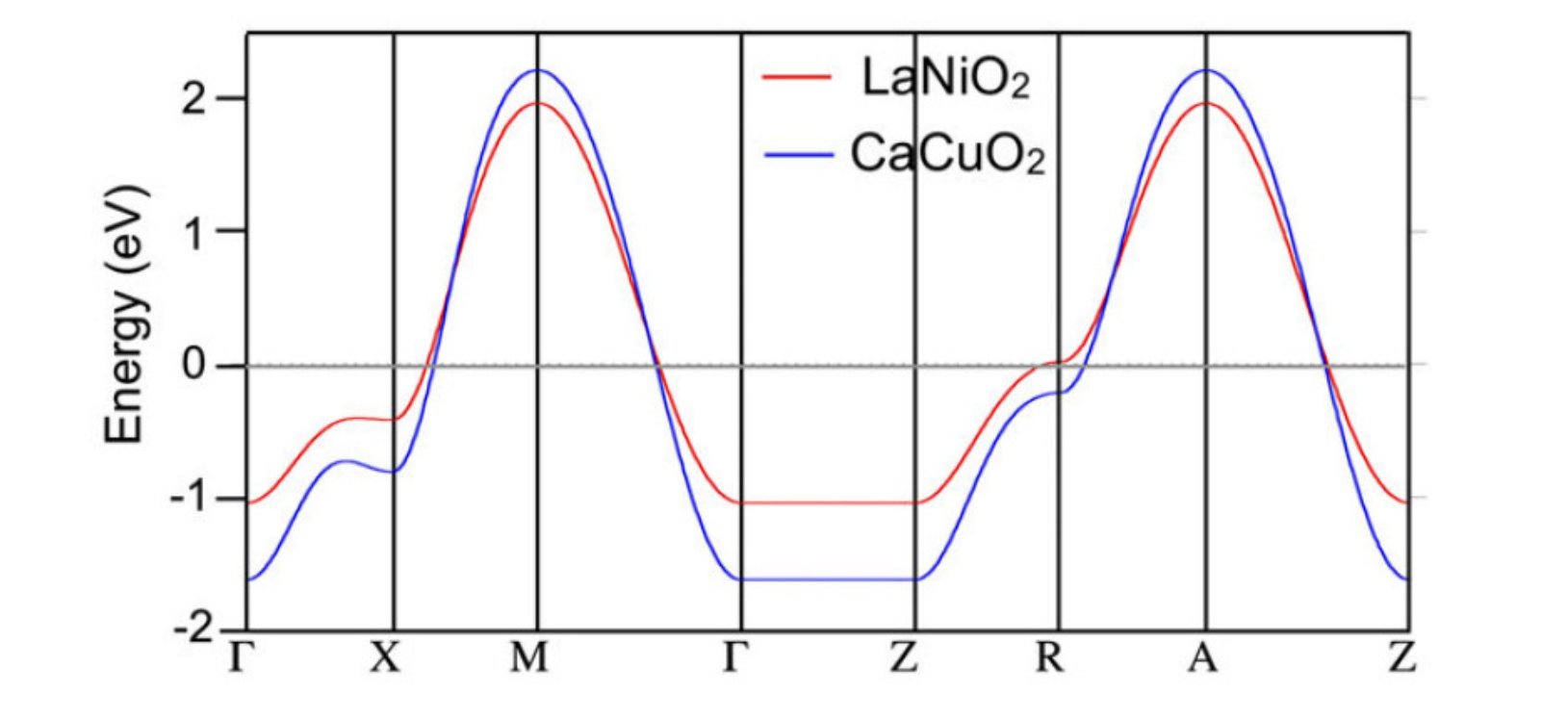}
\caption{
Tight-binding fit to the band with $3d_{x^2-y^2}$ character around the Fermi level for LaNiO$_2$ and CaCuO$_2$. 
Reproduced from Ref.~\cite{Botana_2020}.
NdNiO$_2$ also has smaller bandwidth comprared to CaCuO$_2$ (Fig.~\ref{Fig_NdNiO2_CaCuO2}). 
}
\label{Fig_bandwidth}
\end{center}
\end{figure}

In order to further investigate the similarities and differences between the nickelates and cuprates, 
we show, in Fig.~\ref{Fig_NdNiO2_CaCuO2}, the comparison of the band structures of NdNiO$_2$ and CaCuO$_2$, a copper oxide with the same infinite-layer structure as NdNiO$_2$ (see Refs.~\cite{Lee_2004,Botana_2020} for detail).
The weights of the nickel/copper $3d_{x^2-y^2}$ orbital, the oxygen $2p$ orbitals hybridized with the $3d_{x^2-y^2}$ orbital, and the neodymium/calcium-layer orbitals are also shown as ``fat bands''.

In the case of CaCuO$_2$, the copper $3d$-orbital band is close to the oxygen $2p$-orbital band, and the charge-transfer energy (the potential energy difference between the copper $3d$ orbital and the oxygen $2p$ orbital, which is the energy required for a hole to move from the copper site to the oxygen site) is small.
On the other hand, in the case of NdNiO$_2$, the valence of the nickel cation is about $1+$ and the attraction from the nucleus is small, which lifts the energy level of the $3d$ orbitals and makes the charge transfer energy larger than that of the cuprates.
Due to the higher energy level of the $3d$ orbitals, the $3d_{x^2-y^2}$ band overlaps with the conduction band on the unoccupied side (originating from the Nd-layer orbitals).
The self-doping has thus occurred. 
A larger charge transfer energy makes the hybridization between the nickel $3d_{x^2-y^2}$ and oxygen $2p$ orbitals smaller. 
As a result, the bandwidth of $3d_{x^2-y^2}$ band (more precisely, the antibonding band between the nickel $3d_{x^2-y^2}$ and oxygen $2p$ orbitals) in NdNiO$_2$ is several tens percent smaller than that of CaCuO$_2$ (Fig.~\ref{Fig_bandwidth}).

In Refs.~\cite{Sakakibara_2020,Nomura_2019}, the magnitude of the effective Coulomb interaction for the nickel $3d$ orbital has been estimated, indicating that NdNiO$_2$ is indeed strongly correlated.
However, the presence of the self-doping naturally explains why the parent compound NdNiO$_2$ is not a Mott insulator despite the strong correlation.
Also, the coexistence of the electron and hole carriers suggested by the Hall coefficient measurements is consistent with the multi-band nature around the Fermi level 
(hole carriers from the nickel $3d_{x^2-y^2}$ orbital and electron carriers from the Nd-layer orbitals).

In this section, we mainly discuss the electronic structure of NdNiO$_2$. 
Importantly, a larger charge transfer energy compared to the cuprates and the existence of the self-doping are common features among LaNiO$_2$, PrNiO$_2$, and NdNiO$_2$. 
Indeed, if we assume that the rare-earth $4f$ orbitals are localized and treat them as ``frozen core'', the three compounds show qualitatively similar band structures~\cite{Been_2021,Kapeghian_2020}.  
However, we note that the role of the rare-earth $4f$ orbitals has not been settled; 
A coupling between the rare-earth $4f$ orbitals and the orbitals around the Fermi level may affect the electronic structure around the Fermi level~\cite{P_Jiang_2019,Choi_2020,R_Zhang_2021,Bandyopadhyay_2020}.
We will come back to this point in Sec.~\ref{sec_4f}.

\subsection{Correlation effect on the electronic structure}

Here, we discuss the reconstruction of the electronic structure due to the correlation effects.

\subsubsection{Calculations beyond DFT}
\label{sec_beyond_DFT}

\begin{figure}[tbp]
\vspace{0cm}
\begin{center}
\includegraphics[width=0.48\textwidth]{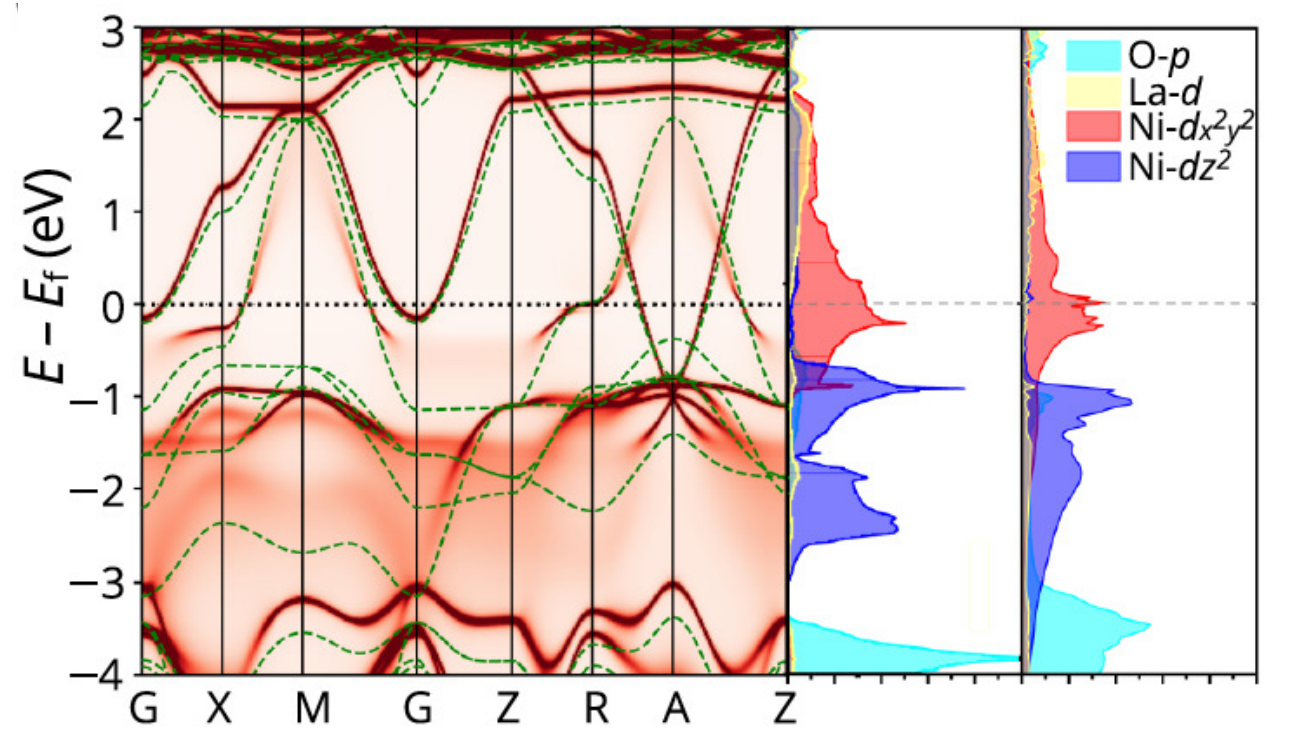}
\caption{
(Leftmost) The DFT+DMFT spectral function for LaNiO$_2$. 
The green dashed curves show the DFT band dispersion. 
(Middle, Rightmost) Orbital-resolved spectral functions from DFT (middle) and DFT+DMFT (rightmost) calculations. 
Reproduced from Ref.~\cite{Ryee_2020}.
}
\label{Fig_DFT+DMFT}
\end{center}
\end{figure}

Many-body effects beyond the DFT level have been investigated using $GW$-type approach~\cite{Hirayama_2020,Olevano_2020,Kutepov_2021}, DMFT(dynamical mean-field theory\cite{Georges_1996,Kotliar_2006})-type approach~\cite{Ryee_2020,Werner_2020,Y_Gu_2020,Liang_Si_2020,Kitatani_2020,Lechermann_2020,Lechermann_2020b,Lechermann_2021,Karp_2020,Karp_2020b,Karp_2021,Leonov_2020,Leonov_2021,X_Wan_2021,Y_Wang_2020,Kang_2021,Z_Liu_2021,Higashi_2021}, and a combination of $GW$- and DMFT-type approaches~\cite{Petocchi_2020,B_Kang_arxiv}.
The main effects of the many-body correlations are the mass renormalization of the correlated orbitals as well as the relative shift of the orbital onsite levels. 
Frequency-dependent self-energy gives rise to incoherent parts in the spectral function (Fig.~\ref{Fig_DFT+DMFT}). 
Whereas the $GW$-type approach is good at describing spatially nonlocal correlations, the DMFT-type approach captures the spatially local correlations well. 
Within the DMFT, the Mott-Hubbard and charge-transfer physics, as well as the local multiplet structure arising from the crystal field and Hund's coupling, can be studied. 

It seems that most of the studies show a qualitative agreement in that the self-doping band robustly remains even with the many-body correlation effects. 
However, unfortunately, so far, there is no consensus on the role of rare-earth-layer orbitals and the hole-doped electron configuration.  
Some works support the presence of Kondo physics arising from the coupling between the itinerant rare-earth-layer orbitals and correlated nickel $3d$ orbitals~\cite{Y_Gu_2020,Lechermann_2020b}.
Some papers propose the importance of Hund's coupling and insist that the multi-orbital nature in the nickel $3d$ manifold is essential~\cite{Werner_2020,Lechermann_2020,Lechermann_2020b,Lechermann_2021,Y_Wang_2020,Z_Liu_2021,Petocchi_2020,B_Kang_arxiv}.
On the other hand, Refs.~\cite{Kitatani_2020,Karp_2020b,Higashi_2021} argue that the important correlation effect lies in the nickel $3d_{x^2-y^2}$ single band.  
The situation might be intermediate between these two~\cite{Kang_2021}. 
Furthermore, Ref.~\cite{Karp_2020} proposes that the charge-transfer physics is important as in the cuprates, while Refs.~\cite{Kitatani_2020,Karp_2020b,Higashi_2021} are based on the Mott-Hubbard-type picture. 
These discrepancies may arise from different conditions (interaction strength, orbital basis, double-counting correction, and so on) employed in the DMFT calculations~\cite{Karp_2021}.
For the controversy on active degrees of freedom, we give an extended discussion in Sec.~\ref{sec_minimal_model}.

\subsubsection{Magnetism}
\label{sec_magnetism}

Another effect of the electron correlation would be inducing some symmetry breaking: magnetism, stripe order, and so on (as for the superconductivity, we discuss in Sec.~\ref{sec_minimal_model}). 
Among them, magnetic instability is of great importance when we compare the infinite-layer nickelates with the cuprates. 

\begin{figure}[tbp]
\vspace{0cm}
\begin{center}
\includegraphics[width=0.48\textwidth]{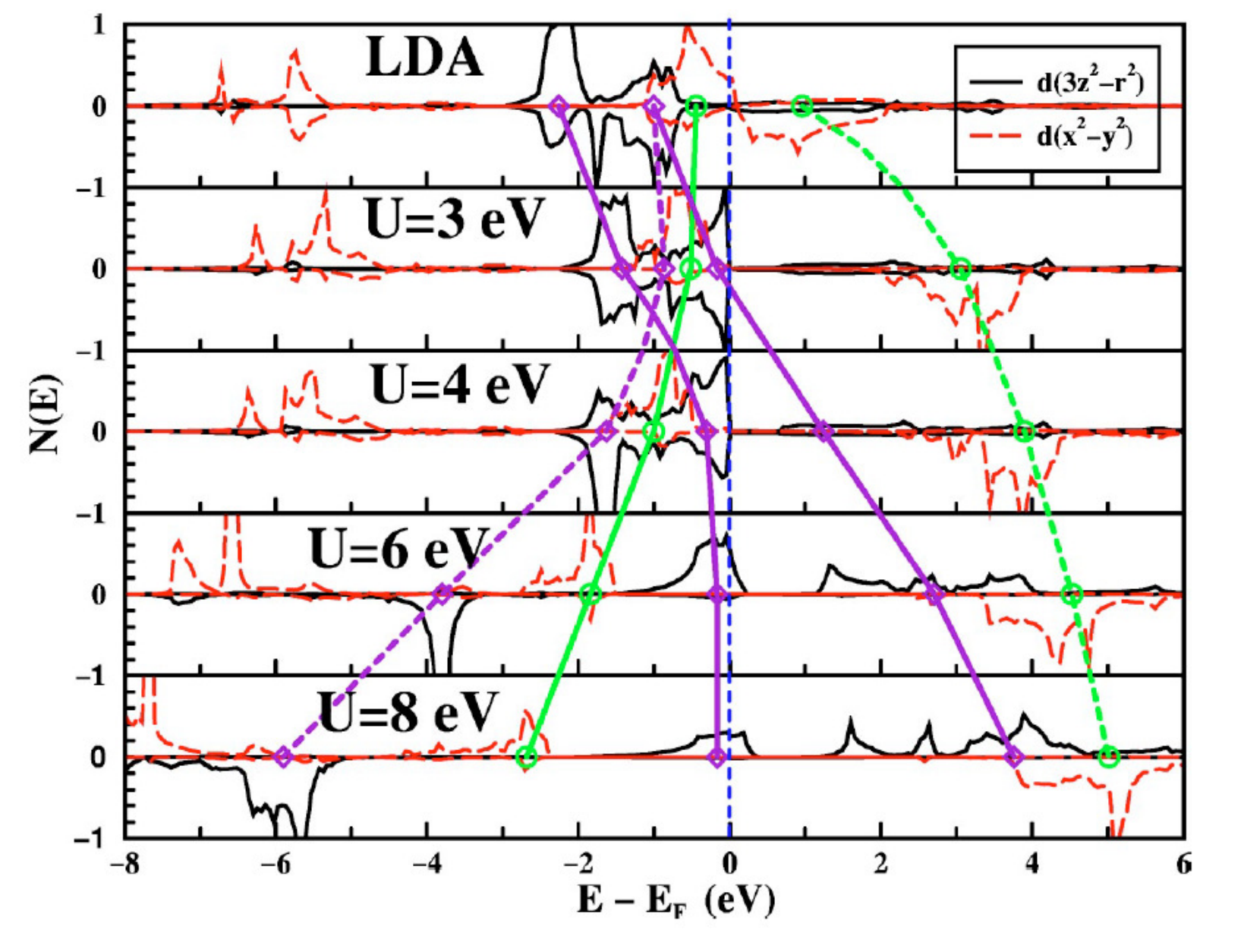}
\caption{
Density of states for nickel $3d_{x^2-y^2}$ and $3d_{3z^2-r^2}$ orbitals in the antiferromagnetic LDA+$U$ calculations for LaNiO$_2$.
The solid and dashed curves outline the path of the density of states of majority and minority spins, respectively, for $3d_{x^2-y^2}$ (green) and $3d_{3z^2-r^2}$ (purple) orbitals.
Reproduced from Ref.~\cite{Lee_2004}.
}
\label{Fig_LSDA}
\end{center}
\end{figure}

So far, there exists no clear evidence for the magnetic long-range order in the parent compounds NdNiO$_2$~\cite{Hayward_2003} and LaNiO$_2$~\cite{Hayward_1999}. 
On the other hand, various theoretical studies have found magnetic solutions~\cite{Lee_2004,Botana_2020,Ryee_2020,HuZhang_2020,Y_Gu_2020,Z_Liu_2020,Lechermann_2021,Karp_2020,Leonov_2020,Leonov_2021,X_Wan_2021,Y_Wang_2020,ZJ_Lang_2021,Choi_2020b} (Fig.~\ref{Fig_LSDA}). 
Although we do not go into detail further because of the lack of experimental evidence for the magnetic order in the parent compounds (for doped bulk Nd$_{0.85}$Sr$_{0.15}$NiO$_2$, an NMR study suggests the presence of short-range antiferromagnetic ordering below 40 K~\cite{Y_Cui_2021}), we emphasize that understanding the magnetism in the nickelates is one of the most urgent and important future tasks. 

Another interesting related issue is the strength of the magnetic exchange coupling $J$, which may be one of the key factors for the high-$T_{\rm c}$ superconductivity~\cite{Lee_Nagaosa_Wen_2006}. 
The cuprates are a charge-transfer material, and $J$ becomes large ($\sim$ 130 meV) due to the superexchange mechanism~\cite{Lee_Nagaosa_Wen_2006}. 

It is an interesting question whether the nickelates, which have larger charge-transfer energy and are more close to the Mott-Hubbard regime, exhibit large $J$ or not. 
The $J$ value for the infinite-layer nickelates is not settled: a Raman experiment using NdNiO$_2$ bulk samples estimated $J$ to be much smaller (25 meV) than that of the cuprates~\cite{Fu_arXiv}; on the other hand, a recent RIXS experiment on NdNiO$_2$ thin film samples gave a larger value of $J = 64(3)$ meV~\cite{Lu_2021}. 
The $J$ value is also scattered in theoretical estimates~\cite{Jiang_2020,Ryee_2020,HuZhang_2020,GM_Zhang_2020,Z_Liu_2020,Been_2021,Leonov_2020,Leonov_2021,X_Wan_2021,ZJ_Lang_2021,Nomura_2020,Katukuri_2020,R_Zhang_2021}. 
One of the reasons for the discrepancy in theoretical estimates is ascribed to the ambiguity in calculating $J$: because the infinite-layer nickelate is not a Mott insulator due to the self-doping, there is ambiguity in mapping the system into the spin models.  
However, there seems to be an overall consensus that the $J$ value is smaller than that of the cuprates. 
The problem is how small it is. 
This is an important question to be clarified in the future because it might be related to the difference in $T_{\rm c}$ between the nickelates and cuprates.

\section{What are essential degrees of freedom for superconductivity?}
\label{sec_minimal_model}

In Sec.~\ref{sec_ele_structure}, we see that, although the self-doping and the large charge-transfer energy make a distinction between the nickelates and cuprates, they are similar in that the strongly correlated $3d_{x^2-y^2}$ orbitals have large Fermi surfaces at the DFT level.
Then, the most interesting and fundamental question would be the pairing mechanism of superconductivity.

Ref.~\cite{Nomura_2019} calculated, from first principles, the electron-phonon coupling constant of NdNiO$_2$ and estimated the transition temperature $T_{\rm c}$ assuming the phonon Bardeen-Cooper-Schrieffer (BCS)~\cite{Bardeen_1957} mechanism. 
The estimated $T_{\rm c}$ is less than 1 K.
Thus, the phonon mechanism cannot explain the experimental $T_{\rm c}$ on the order of 10 K. 
This suggests that the superconductivity in doped NdNiO$_2$ originates from an unconventional mechanism (we note that the electron-phonon interactions may contribute to the superconductivity in cooperation with other mechanisms).
Indeed, unconventional mechanisms have been proposed theoretically from the early stage~\cite{Sakakibara_2020,Hirsch_2019,Wu_2020}.
Experimentally, the STM/STS experiment has observed a $d$-wave-like gap in some regions of the inhomogeneous surfaces of doped NdNiO$_2$ films~\cite{Q_Gu_2020} (see also Sec.~\ref{sec_experiment_indivisual}).

When discussing unconventional mechanisms, it is helpful to construct an effective lattice Hamiltonian, such as the Hubbard model, $t$-$J$ model, $d$-$p$ model, or periodic Anderson model, for the electronic degrees of freedom near the Fermi level and analyze the superconductivity based on it.
The effective Hamiltonian should reflect the crystal and electronic structures of the system. 
Then, what is the minimum model to describe the superconductivity in the infinite-layer nickelates (different models incorporate different physics such as Mott, Hund, and Kondo physics, as we discuss below)? 
In other words, what are the essential degrees of freedom for describing superconductivity? 
In this section, we will discuss this question.

So far, many different proposals have been made. 
However, there seems to be a consensus that the nickel $3d_{x^2-y^2}$ orbital is one of the essential degrees of freedom.
Candidates for other key degrees of freedom include 
\begin{enumerate}
    \item the itinerant rare-earth-layer orbitals that form the additional Fermi pockets
    \item the rare-earth $4f$ orbitals
    \item the oxygen $2p$ orbitals that hybridize with $3d_{x^2-y^2}$ orbitals on the NiO$_2$ plane ($p_x$ orbital of the oxygen in the $x$-direction and $p_y$ orbital of oxygen in the $y$-direction)
    \item nickel $3d$ orbitals other than the $3d_{x^2-y^2}$ orbital
\end{enumerate}
Each of these is discussed in the following.

\subsection{Itinerant rare-earth-layer orbitals forming additional Fermi pockets}

As we see in Sec.~\ref{sec_ele_structure}, the rare-earth-layer orbitals form Fermi pockets around the $\Gamma$ and $A$ points. The X-ray experiments on film samples consistently suggest that the rare-earth-layer orbitals are partially occupied~\cite{Hepting_2020}. 
Therefore, when discussing the symmetry of the superconductivity gap, we need to consider the gap functions on these Fermi surfaces.
A key question, however, is whether the superconducting gaps on these bands are a byproduct of the superconductivity in the $3d_{x^2-y^2}$ band or whether they play an intrinsic role in the emergence of superconductivity.
To put it a little differently, in discussing the properties such as superconductivity and magnetism, are the rare-earth-layer orbitals more than ``charge reservoir'' controlling the filling of the nickel $3d_{x^2-y^2}$ orbitals?

\begin{figure}[tbp]
\vspace{0cm}
\begin{center}
\includegraphics[width=0.48\textwidth]{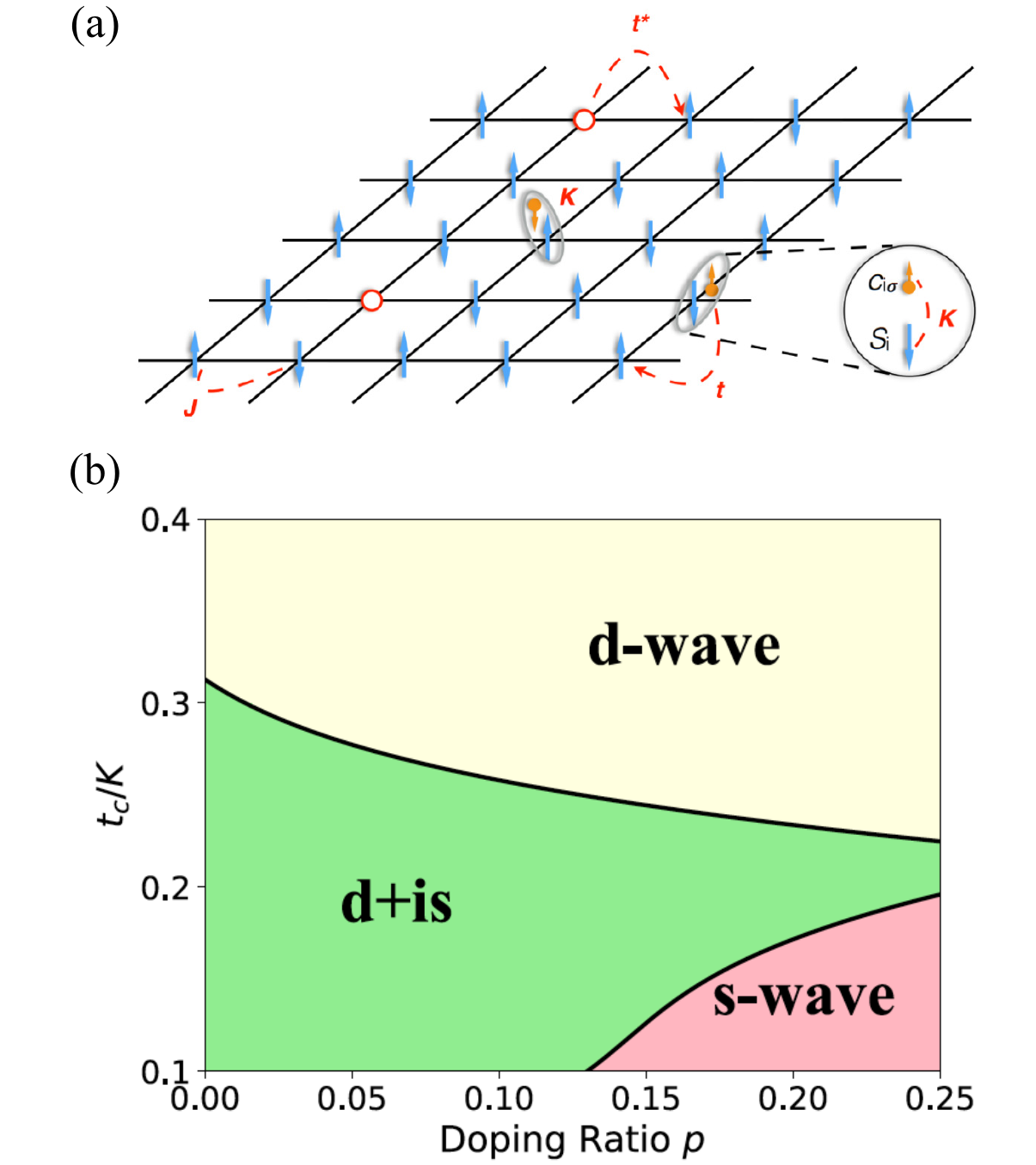}
\caption{
(a) Effective two-dimensional model on the square lattice proposed in Ref.~\cite{GM_Zhang_2020}.
Blue and orange arrows represent the nickel spins and rare-earth $5d$ electrons, respectively. 
$J$ and $K$ are the magnetic exchange coupling between the neighboring nickel spins and the Kondo coupling between the nickel spins and rare-earth $5d$ electrons, respectively. 
Red circles indicate nickel $d^8$ configuration (holon). $t$ and $t^*$ are the hopping integrals of doublons and holons, respectively.
Reproduced from Ref.~\cite{GM_Zhang_2020}. 
(b) Phase diagram of superconductivity obtained by generalized $K$-$t$-$J$ model, where the Kondo coupling $K$ to the itinerant electrons are added to the $t$-$J$ model.
$t_{\rm c}$ is the conduction electron hopping. 
The parameters in the $t$-$J$ part is taken to be $t/K =0.2$, $t'/K = -0.05$, $J/K=0.1$, where $t$, $t'$, and $J$ are nearest-neighbor hopping, next-nearest-neighbor hopping, and exchange coupling, respectively. 
Reproduced from Ref.~\cite{Z_Wang_2020}.
}
\label{Fig_Kondo}
\end{center}
\end{figure}

If the hybridization between the nickel $3d$ and rare-earth-layer orbitals is substantial, the rare-earth-layer orbitals are not only a charge reservoir, but they might give Kondo-like physics~\cite{Sawatzky_2019,GM_Zhang_2020}:
The rare-earth-layer electrons have large bandwidth and couple to localized spins at nickel sites as itinerant conduction electrons. 
In this case, the increase in the electrical resistivity seen in NdNiO$_2$ from around 70 K~\cite{Li_2019} is interpreted as the Kondo effect~\cite{Sawatzky_2019,GM_Zhang_2020}.
Following this picture, effective Hamiltonians that include both nickel $3d$ and rare-earth-layer orbitals have been proposed~\cite{Hepting_2020,GM_Zhang_2020,Y_Gu_2020}.
Ref.~\cite{Z_Wang_2020} has analyzed superconductivity based on the model proposed in Ref.~\cite{GM_Zhang_2020} and shown pairing instabilities towards $d$, $d+is$, and $s$ waves depending on the parameter region (Fig.~\ref{Fig_Kondo}).

The presence of Kondo physics is under debate: 
the recent experiments on the resistivity of magnetic-field-induced normal state using Nd$_{1-x}$Sr$_{x}$NiO$_2$ film samples have reported the resilience of the resistivity upturn against magnetic field and the positive magnetoresistance~\cite{Hsu_2021}.
Ref.~\cite{Hsu_2021} poses a question on the Kondo scenario, whose effect is expected to be suppressed by the magnetic field. 
On the other hand, Ref.~\cite{Ikeda_2016} observes negative magnetoresistance for LaNiO$_2$ film samples. 
The Kondo temperature is affected by the density of states and Fermi energy of the conduction bands and the strength of the Kondo coupling between the local spins and itinerant electrons~\cite{Hewson_1993}.
The density of states of the rare-earth-layer orbital is not large~\cite{Nomura_2019}. 
As for the hybridization between the nickel $3d_{x^2-y^2}$ and rare-earth layer orbitals giving the Kondo coupling, some works argue that the hybridization is weak~\cite{Kitatani_2020,Karp_2020}, whereas others emphasize its importance~\cite{Y_Gu_2020,Lechermann_2020b}.

\begin{figure*}[tbp]
\vspace{0cm}
\begin{center}
\includegraphics[width=0.96\textwidth]{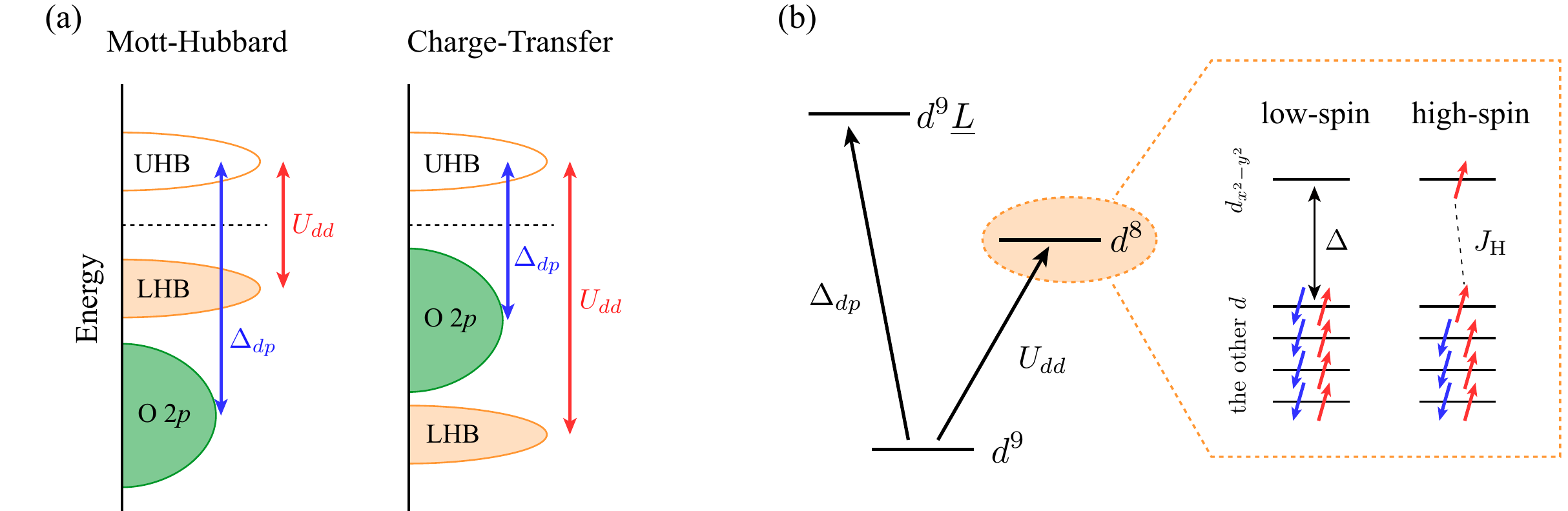}
\caption{
(a) Schematic figure showing the difference between a Mott-Hubbard insulator and a charge-transfer insulator.
UHB and LHB are the upper Hubbard and lower Hubbard bands of the $3d$ orbitals, respectively.
(b) Schematic energy diagram for the hole doping. 
There is a large energy scale competition between $d^8$ and $d^9 \underline{L}$ ($U_{dd}$ vs $\Delta_{dp}$). 
Within the $d^8$ configuration, a smaller energy scale competition exists between the low-spin ($S=0$) and high-spin ($S=1$) states  ($\Delta$ vs $J_{\rm H}$). 
$\Delta$ is the crystal-field splitting between the $3d_{x^2-y^2}$ orbital and the other $3d$ orbitals [see Fig.~\ref{Fig_RNiO2}(b)]. 
$J_{{\rm H}}$ is Hund's coupling of the nickel $3d$ orbitals.
}
\label{Fig_energy_diagram}
\end{center}
\end{figure*}

\subsection{Rare-earth 4f orbitals}
\label{sec_4f}

At the moment, the role of the rare-earth $4f$ orbitals is an open question. 
Since the $4f$ orbitals are spatially localized, the bandwidth of $4f$ orbitals becomes small.
Because of the correlation effect, the $4f$ electrons tend to be localized and form local magnetic moments. 
The disordered localized spins on the rare-earth layer may affect the transport property on the NiO$_2$ plane. 
Theoretically, the role of $4f$ orbitals has been investigated using the DFT-based calculations~\cite{P_Jiang_2019,Choi_2020,R_Zhang_2021,Bandyopadhyay_2020} and 
DFT+DMFT~\cite{Z_Liu_2021}. 
In particular, the DFT-based calculations suggest that an intraatomic interaction/hybridization between the rare-earth $4f$ and $5d$ orbitals may affect the energy level of the self-doping band~\cite{Choi_2020,Bandyopadhyay_2020}, thus changing the Fermi pocket size. 
A hybridization between the rare-earth $4f$ and nickel $3d$ orbitals is also discussed~\cite{P_Jiang_2019,R_Zhang_2021}.
On the other hand, the DFT+DMFT results show that the bands around the Fermi level are almost unaffected by the rare-earth $4f$ orbitals~\cite{Z_Liu_2021}. 
Experimentally, on top of the superconductivity in doped NdNiO$_2$ and PrNiO$_2$ films with different $4f$ occupations, the superconductivity has been recently found in doped LaNiO$_2$~\cite{Osada_2021,SW_Zeng_2022} film samples, where the $4f$ orbitals are empty. 
The common observation of superconductivity irrespective of the filling of $4f$ orbitals would imply that the $4f$ degrees of freedom are not the main player in realizing superconductivity.

\subsection{Oxygen 2p orbitals}

\begin{figure}[tbp]
\vspace{0cm}
\begin{center}
\includegraphics[width=0.48\textwidth]{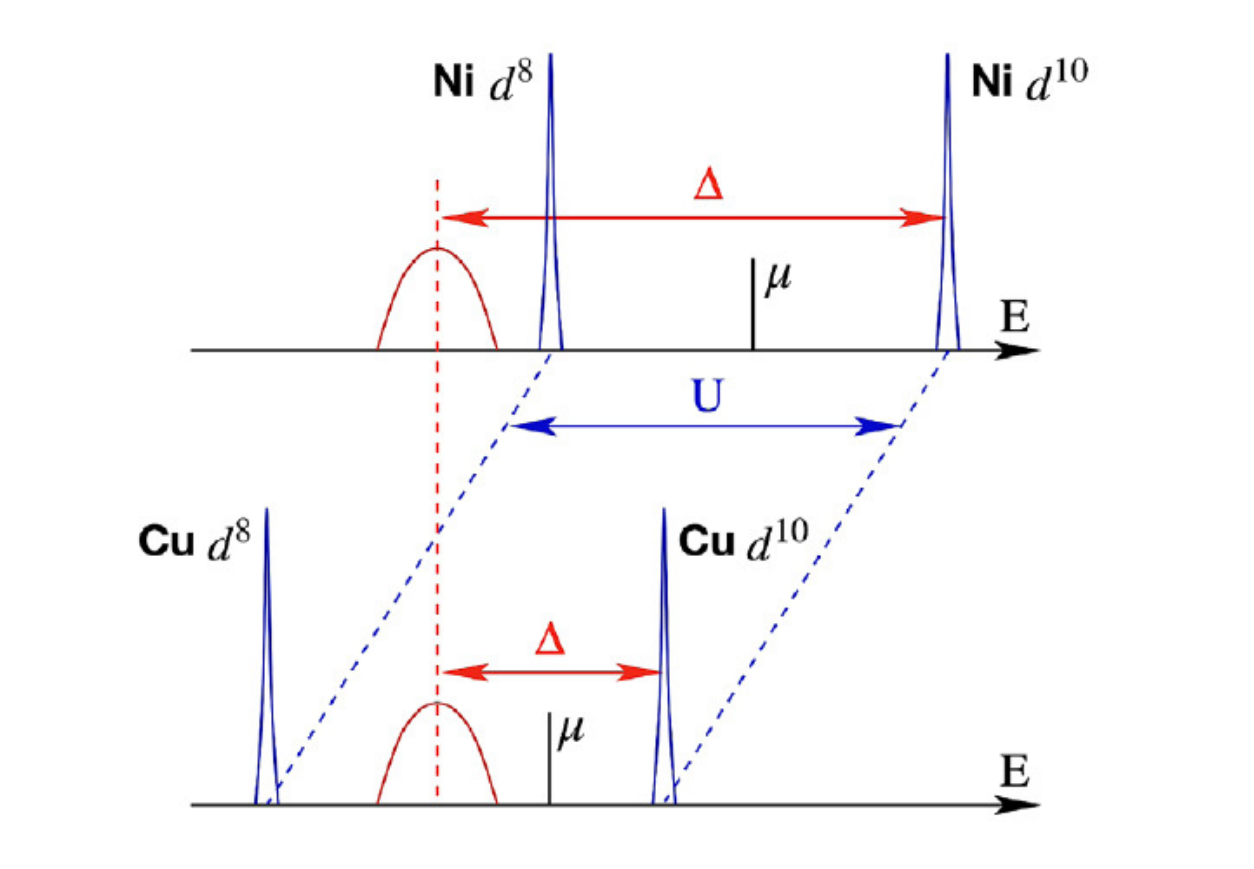}
\caption{
Schematic figure comparing the cuprates and nickelates assuming the charge-transfer energy $\Delta_{dp}$ (denoted as $\Delta$ in the figure) 
in the nickelates is much larger than that of the cuprates. 
The blue bands show Hubbard bands, while the red bands are oxygen $2p$ band with the $pd$ hybridization being switched off.
Reproduced from Ref.~\cite{Jiang_2020}.
}
\label{Fig_Mott-Hubbard}
\end{center}
\end{figure}

It is of great importance to consider the role of oxygen $2p$ orbitals when comparing the cuprates and nickelates. 
A problem is to which orbitals the doped holes go (see Figs.~\ref{Fig_energy_diagram} and \ref{Fig_Mott-Hubbard}). 
In the cuprates classified as a charge-transfer insulator in the Zaanen-Sawatzky-Allen phase diagram \cite{Zaanen_1985}, the doped holes mainly enter the oxygen $2p$ orbitals because the charge-transfer energy $\Delta_{dp}$ is smaller than the $3d$-orbital Hubbard interaction $U_{dd}$.
Therefore, the electron configuration of the hole-doped cuprates is mainly $d^9 \underline{L}$, 
and the physics of Zhang-Rice singlet emerges~\cite{Zhang_1988}.

On the other hand, in the case of the infinite-layer nickelates $R$NiO$_2$, $\Delta_{dp}$ is larger than that of the cuprates (see Sec.~\ref{sec_ele_structure}).
This suggests that $R$NiO$_2$ is a Mott-Hubbard type material.  
The doped holes mainly go to the nickel $3d$ orbitals, and the doped electron configuration is mainly represented by $d^8$ (doping is expected to make the rare-earth-layer Fermi pockets smaller and weaken the self-doping effect).
Note, however, that the hybridization between the nickel $3d_{x^2-y^2}$ orbital and the oxygen $2p$ orbitals is nonzero. 
Therefore, some holes should go to oxygen $2p$ orbitals~\cite{Karp_2020,ZJ_Lang_2021}. 
In fact, the O K-edge EELS measurement has detected a signal of holes in the oxygen $2p$ orbitals, but the intensity is much smaller than that of the cuprates~\cite{Goodge_2021} (see also Sec. \ref{sec_experiment_indivisual}).

\subsection{Other 3d orbitals}
\label{sec_other_3d}

Based on the above discussion, the undoped $d^9$ configuration and the $d^8$ configuration with holes in the nickel $3d$ orbitals would be important in describing the superconductivity in the infinite-layer nickelates. 
When we follow this picture, an effective Hamiltonian would consist of nickel $3d$ orbitals
(note that, in this case, the hybridization of the oxygen $2p$ orbitals with the nickel $3d_{x^2-y^2}$ orbital is taken into account by considering $3d_{x^2-y^2}$-like Wannier orbitals centered at nickel sites with oxygen $2p$ tails).
This raises a question: which model is more appropriate, the single-orbital or multi-orbital $3d$ model?

\begin{figure}[tbp]
\vspace{0cm}
\begin{center}
\includegraphics[width=0.48\textwidth]{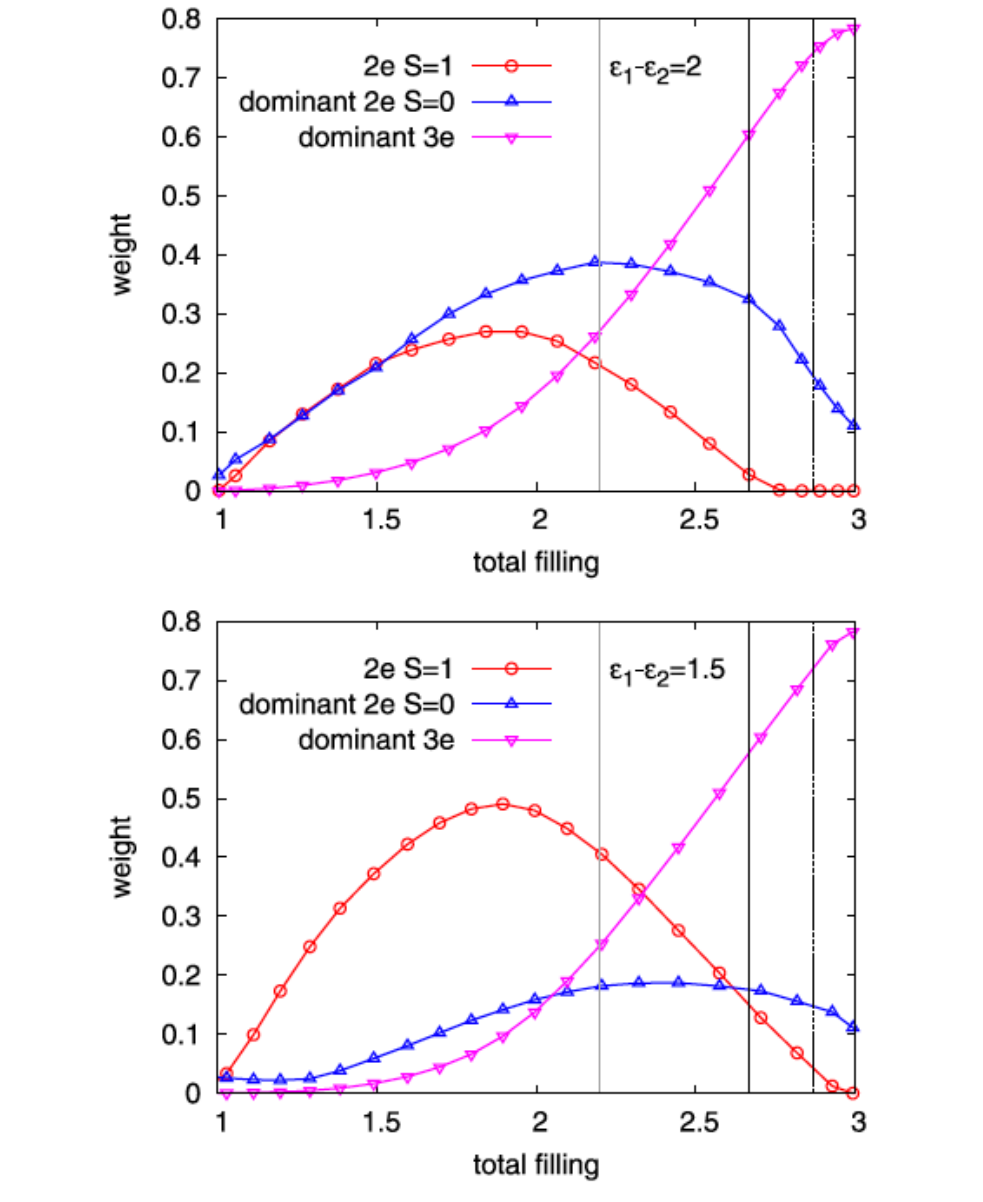}
\caption{
The DMFT results for the probability of local electron configurations for a two-orbital model mimicking nickel $3d_{x^2-y^2}$ and $3d_{3z^2-r^2}$ orbitals. 
The noninteracting density of states for these two orbitals is approximated by the semicircular density of states with the bandwidth of 3 ($3d_{x^2-y^2}$) and 2 ($3d_{3z^2-r^2}$).
The Slater-Kanamori type interaction with $U=2.6$, $U'=1.3$, and $J=0.5$ is employed with the Hubbard interaction $U$, the interorbital interaction $U'$, and the Hund's coupling $J$. 
$\epsilon_1$ and $\epsilon_2$ are the onsite level of $3d_{x^2-y^2}$ and $3d_{3z^2-r^2}$ orbitals, respectively. 
Reproduced from Ref.~\cite{Werner_2020}.
}
\label{Fig_multi-orbital}
\end{center}
\end{figure}

This can be understood as a competition between Hund's coupling and the crystal-field splitting, 
where the former (latter) favors the high-spin (low-spin) configuration [Fig.~\ref{Fig_energy_diagram}(b)].
If the crystal-field splitting between the $3d_{x^2-y^2}$ orbital and the other $3d$ orbitals is sufficiently large, the holes stay in the $3d_{x^2-y^2}$ orbitals. 
Then, the doped $d^8$ configuration takes the low-spin state ($S\!=\!0$), and the single-orbital picture would be justified.
If Hund's coupling induces the high-spin ($S\!=\!1$) $d^8$ configuration, the multi-orbital model becomes indispensable.
Several theoretical studies have argued that the multi-orbital nature cannot be ignored~\cite{Jiang_2020,YH_Zhang_2020,Werner_2020,Petocchi_2020,LH_Hu_2019,Lechermann_2020,Lechermann_2020b,Lechermann_2021,J_Chang_2020,Y_Wang_2020,Z_Liu_2021,B_Kang_arxiv,Choi_2020b} (Fig.~\ref{Fig_multi-orbital}).
On the other hand, Refs.~\cite{Kitatani_2020,Karp_2020b,Higashi_2021} argue a single-orbital picture (for example, a recent DFT+DMFT study~\cite{Higashi_2021} has well reproduced the XAS, XPS, and RIXS spectra of NdNiO$_2$ in Refs.~\cite{Hepting_2020,Fu_arXiv,Rossi_2021} with the low-spin ground state). 
An intermediate picture also exists: while the low-spin state is realized at low energy, the trace of Hund's coupling can be observed as dynamical orbital fluctuations at high frequencies~\cite{Kang_2021}.
Experimentally, Ref.~\cite{Rossi_2021} performed RIXS and XAS measurements using the doped NdNiO$_2$ film samples and suggested a low-spin character of the doped configuration
(see also Ref.~\cite{Matsumoto_2019} and references therein for the situations for other $d^8$ nickelates with the square-planar environment).

\begin{figure}[tbp]
\vspace{0cm}
\begin{center}
\includegraphics[width=0.51\textwidth]{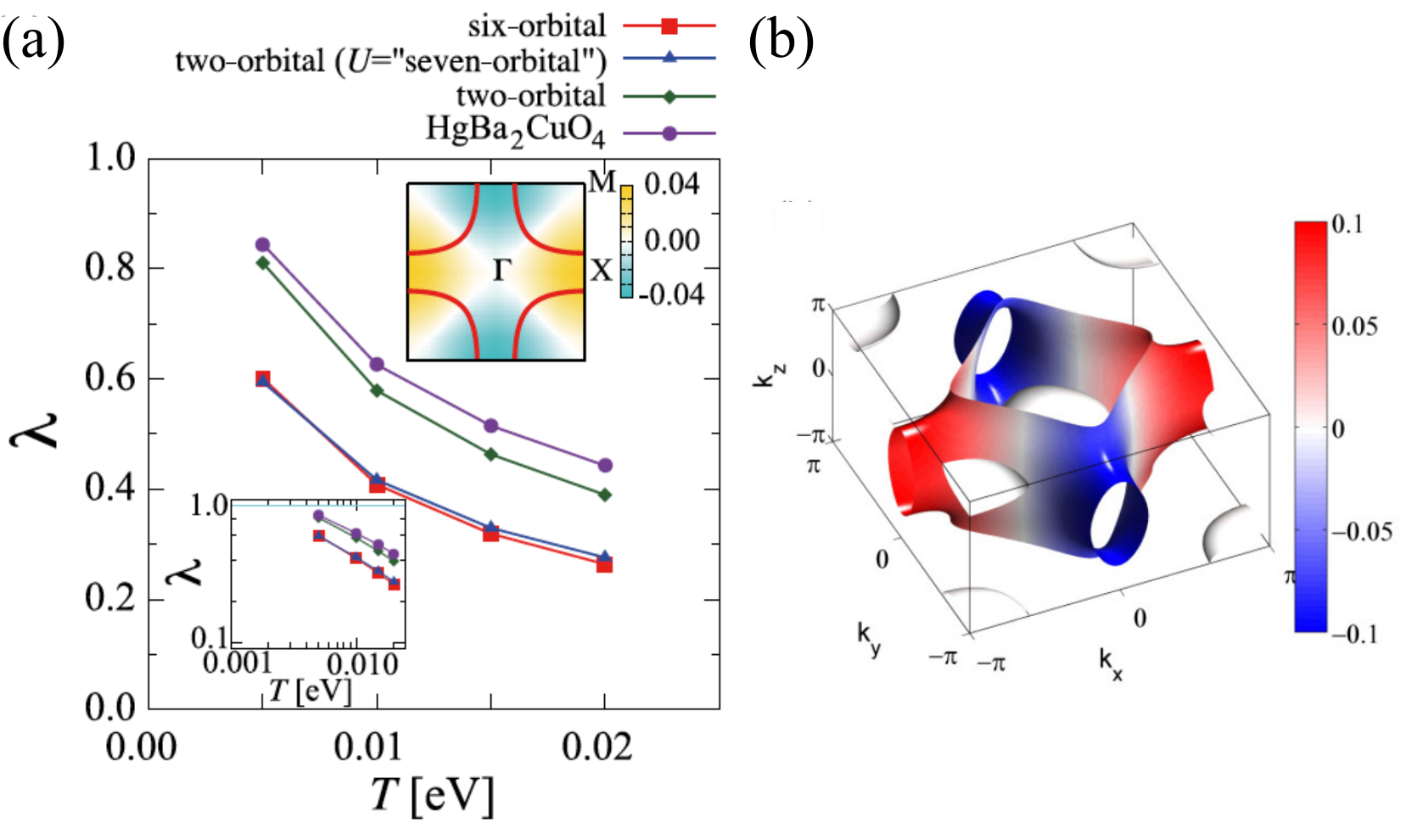}
\caption{
(a) Temperature evolution of the $d_{x^2-y^2}$-wave pairing instability at 20 \% hole doping for effective models of LaNiO$_2$.
$\lambda$ is the eigenvalue of the Eliashberg equation. 
Three different models for LaNiO$_2$ are compared with the five orbital model of HgBa$_2$CuO$_4$ (see Ref.~\cite{Sakakibara_2020}  for details).
Ref.~\cite{Sakakibara_2020} argues that the superconducting instability in the nickelates is weaker than that of the cuprates because of a larger interaction and the resulting self-energy renormalization effect.  
The insets show $\lambda$-$T$ log-log plot (bottom left) and the eigenfunction of the Eliashberg equation at $k_z=0$ (top right). 
Reproduced from Ref.~\cite{Sakakibara_2020}. 
(b) The superconducting gap of $d_{x^2-y^2}$-wave pairing at 20 \% hole doping in the $t$-$J$ model.
Reproduced from Ref.~\cite{Wu_2020}.
}
\label{Fig_d-wave}
\end{center}
\end{figure}

\subsection{Brief summary}

\begin{figure}[tbp]
\vspace{0cm}
\begin{center}
\includegraphics[width=0.48\textwidth]{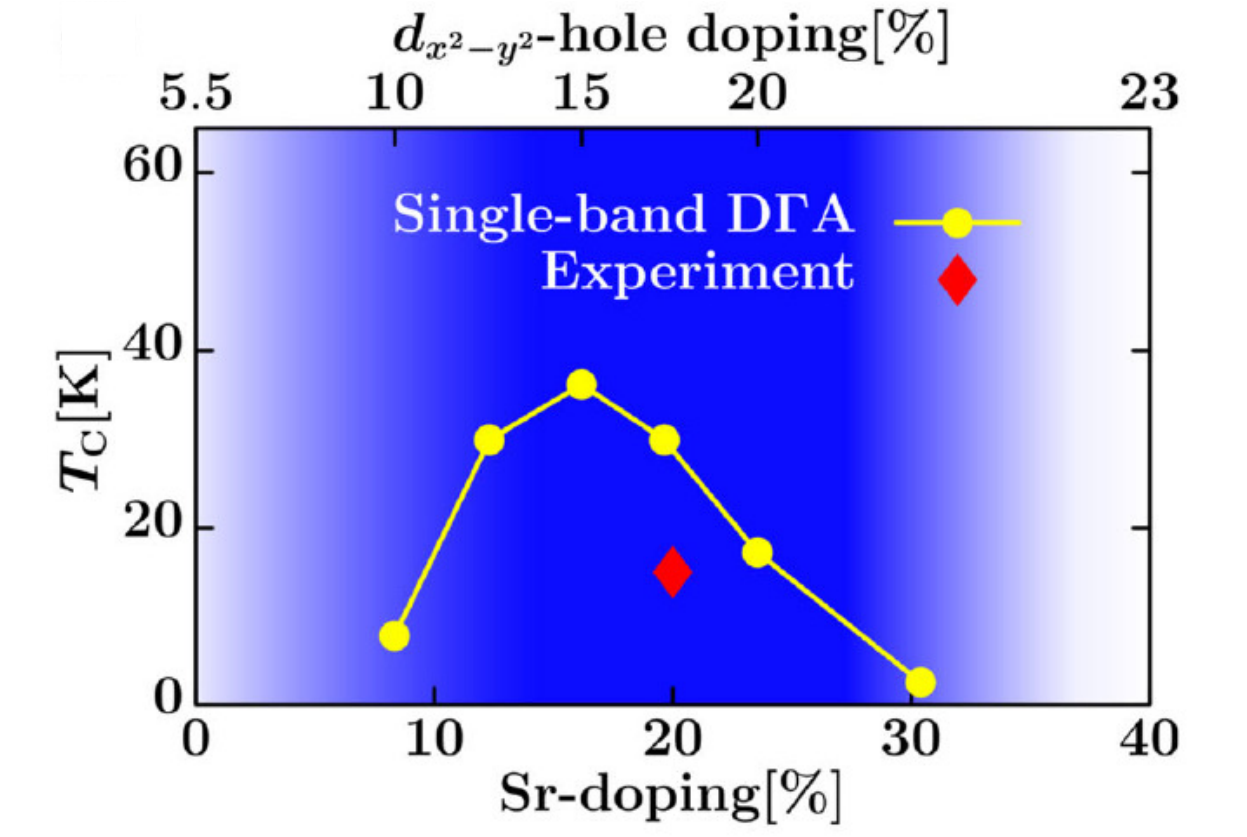}
\caption{
The superconducting transition temperature $T_{\rm c}$ of the $d$-wave pairing as a function of Sr-doping. 
The calculation is performed for a single-orbital Hubbard model using the dynamical vertex approximation (D$\Gamma$A). 
In the blue-shaded region, a single-orbital Hubbard model description is argued to be possible.    
Reproduced from Ref.~\cite{Kitatani_2020}.
}
\label{Fig_single-orbital}
\end{center}
\end{figure}

To summarize the above discussion, most of the works view that the correlation effects in the nickel $3d_{x^2-y^2}$ orbital are important. Indeed, {\it ab initio} estimates of the effective Coulomb interactions within the nickel $3d$ manifold have suggested that the infinite-layer nickelates are strongly correlated systems~\cite{Sakakibara_2020,Nomura_2019}.
When one considers the nickel $3d_{x^2-y^2}$ orbital to be the most essential degrees of freedom, the $d$-wave pairing might be plausible: 
theoretically, spin-fluctuation-induced $d$-wave pairing for the nickel $3d_{x^2-y^2}$ orbital has been proposed \cite{Sakakibara_2020,Wu_2020} (Fig.~\ref{Fig_d-wave}).
From the standpoint of the single-orbital picture, the experimentally observed dome-like $T_{\rm c}$ \cite{D_Li_2020,Zeng_2020,Osada_2020_2,Osada_2021,SW_Zeng_2022} would be related to the change in superconducting instability as a function of the occupation of the nickel $3d_{x^2-y^2}$ orbital~\cite{Kitatani_2020} (Fig.~\ref{Fig_single-orbital}).

On the other hand, if we take a position that the multi-orbital nature is important, a comparison with iron-based superconductors may be interesting~\cite{Y_Wang_2020}.
Enhanced dynamical spin fluctuations around spin-freezing (a phenomenon where the local moment is very slowly fluctuating, i.e., almost frozen) crossover induced by physical and effective Hund's coupling may be relevant to the superconductivity~\cite{Werner_2020}.
Strong-coupling-expansion type approach using the $3d$ multi-orbital model may help understanding the unconventional pairing~\cite{YH_Zhang_2020,LH_Hu_2019,J_Chang_2020}. Even when the Fermi surfaces of the other $3d$ orbitals do not exist, if other $3d$ bands exist just below the Fermi level (incipient band), such incipient bands can play a role in enhancing the superconductivity~\cite{Kitamine_2020}.

When the rare-earth-layer orbitals are active, they lead to another type of unconventional pairing through the Kondo coupling~\cite{Z_Wang_2020}.
Even when they are not the main player in the development of superconductivity, they do form Fermi pockets. 
Thus, the superconducting gap should also be finite on these additional bands. 
The interactions between the nickel $3d$ orbitals and the rare-earth-layer orbitals would be an important factor in determining a gap structure for the additional Fermi pockets~\cite{Adhikary_2020}.
It is an important future task for experiments to reveal the gap structure on multiple Fermi pockets 
(a starting point is given by Ref.~\cite{Q_Gu_2020}, see Sec.~\ref{sec_experiment_indivisual}).

\section{Searching for a new addition to the family}
\label{sec_new_materials}

As we describe in Sec.~\ref{sec_experiment}, rare-earth infinite-layer nickelates show superconductivity for rare-earth elements of La, Pr, and Nd.
Superconductivity may also appear for other rare-earth element cases, which is an interesting future task to be elucidated. 
Another intriguing issue is whether the superconductivity appears only in the infinite-layer structure or not. 
Here, we discuss several possible candidates for a new member of the family of nickel-based superconductors.

\subsection{Multi-layer nickelates}
\label{Sec:multi-layer}

The most natural extension of the infinite-layer nickelates would be the multi-layer square-planar nickelates~\cite{Nica_2020,Botana_2021}. 
This is because the rare-earth infinite-layer nickelates are a special case ($n=\infty$) of the multi-layer square-planar nickelates $R_{n+1}$Ni$_n$O$_{2n+2}$.
In the case of the cuprates, superconductivity appears both in infinite-layer and multi-layer structures \cite{note_Mukuda_2012}.

\begin{figure}[tbp]
\vspace{0cm}
\begin{center}
\includegraphics[width=0.48\textwidth]{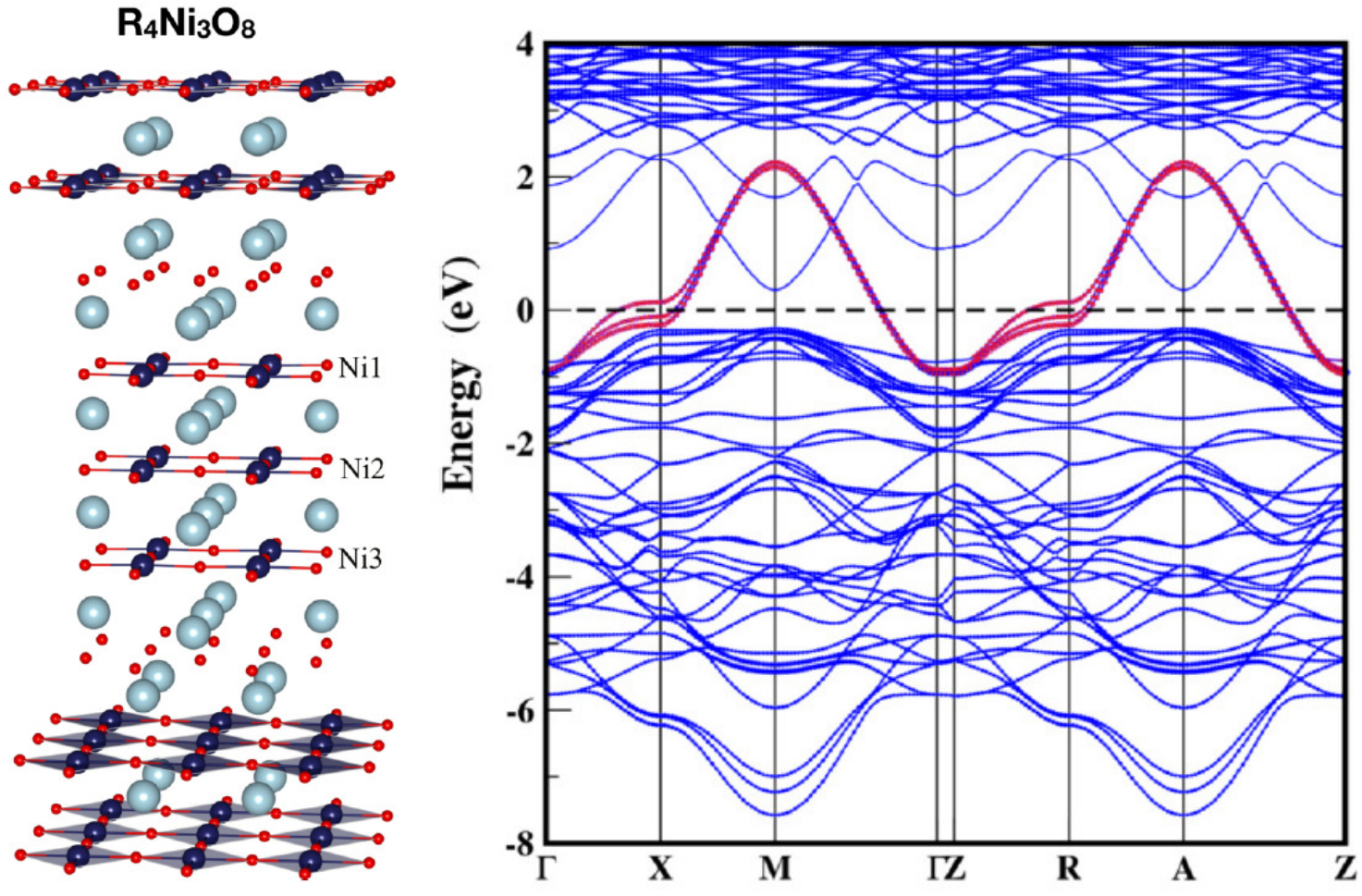}
\caption{
The crystal structure (left) and the DFT band dispersion (right) of a trilayer nickelate $R_4$Ni$_3$O$_8$.  
In the DFT calculation, $R$=La is employed. 
The bands with dominant nickel $3d_{x^2-y^2}$-orbital character are highlighted in red. 
Reproduced from Ref.~\cite{Nica_2020}.
}
\label{Fig_438}
\end{center}
\end{figure}

\begin{figure*}[tbp]
\vspace{0cm}
\begin{center}
\includegraphics[width=0.95\textwidth]{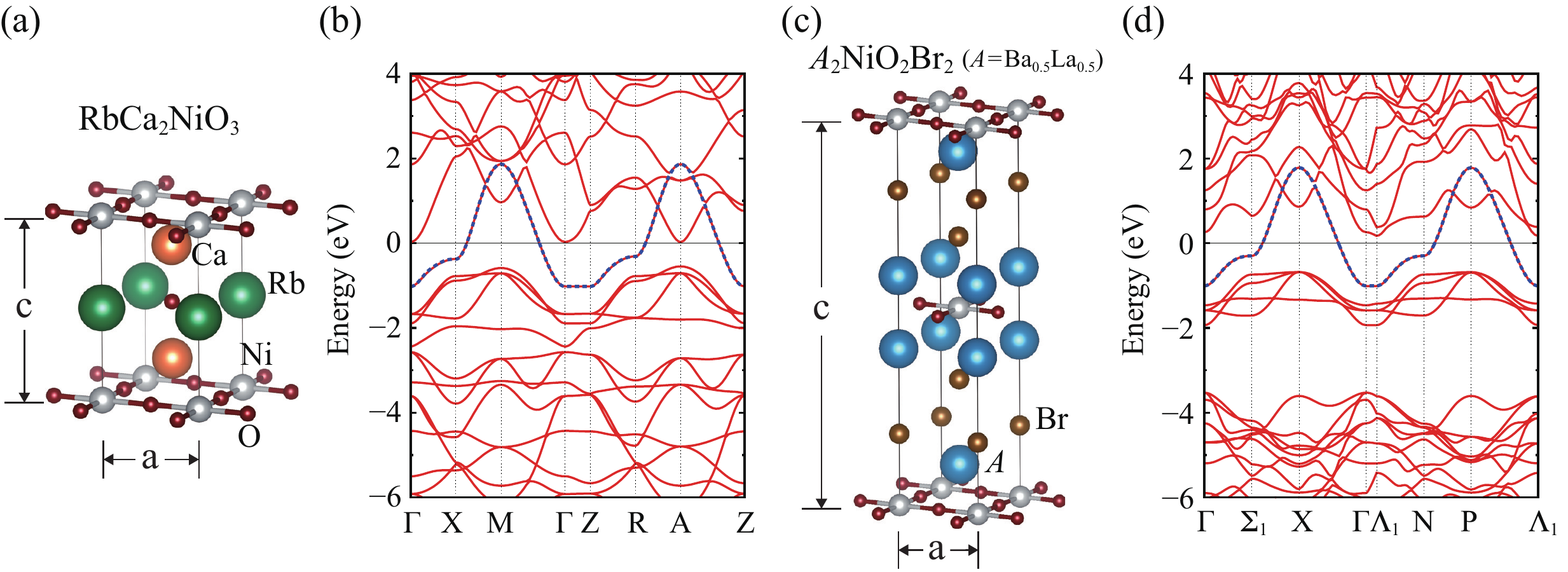}
\caption{
Two $d^9$ layered nickelates without self-doping proposed in Ref.~\cite{Hirayama_2020}.
Crystal structures of (a) RbCa$_2$NiO$_3$ and (c) $A_2$NiO$_2$Br$_2$ ($A$ is a $2.5+$-valent cation, here Ba$_{0.5}$La$_{0.5}$ is assumed), and (b,d) the corresponding DFT-GGA band structures.
In (b) and (d), the ${\bf k}$ path is consistent: $(0,0,0)$ $\rightarrow$ $(\pi/a,0,0)$ $\rightarrow$ $(\pi/a,\pi/a,0)$ $\rightarrow$  $(0,0,0)$ $\rightarrow$ $(0,0,\pi/c)$ $\rightarrow$ $(\pi/a,0,\pi/c)$ $\rightarrow$ $(\pi/a,\pi/a,\pi/c)$ $\rightarrow$  $(0,0,\pi/c)$ (The symbols are different because the primitive cells of RbCa$_2$NiO$_3$ and $A_2$NiO$_2$Br$_2$ are tetragonal and bace-centered tetragonal, respectively).
The blue dotted curves are the Wannier band dispersion of an effective single-orbital model of the nickel $3d_{x^2-y^2}$ Wannier orbital with the oxygen-$2p$ tails. 
The band structure is metallic in the DFT-GGA calculation. 
However, they are expected to be Mott insulators if we properly take into account the correlation effects. 
Reproduced from Ref.~\cite{Nomura_2020}.
}
\label{Fig_d9_nickel}
\end{center}
\end{figure*}

Multi-layer square-planar nickelates with $n=2$ and $3$ had already been synthesized using the reduction process from the Ruddlesden-Popper parent compound $R_{n+1}$Ni$_n$O$_{3n+1}$ (see, e.g., Ref.~\cite{J_Zhang_2021} for a review).
This situation is similar to the infinite-layer case, where $R$NiO$_2$ is obtained by the reduction from $R$NiO$_3$. 
An average nominal valence of the nickel ions in the $n=3$ compound is $1.33+$ (i.e., $d^{8.67}$), which corresponds to the overdoped regime in the cuprates (see Fig.~\ref{Fig_438} for the crystal structure and the DFT band dispersion).  
Indeed, trilayer nickelates ($n=3$) do not show superconductivity~\cite{J_Zhang_2017}.
However, there is a signature for a large magnetic exchange coupling about 70 meV in the trilayer nickelate La$_4$Ni$_3$O$_8$, which makes multi-layer nickelates interesting reference systems between the cuprates and infinite-layer nickelates~\cite{JQ_Lin_2021}. 
Increasing $n$ would put the filling of nickel $3d$ orbital of multi-layer nickelates in the right place for possibly realizing superconductivity~\cite{Norman_2020,Botana_2021}.
In this respect, a notable experimental advance is a recent report of the epitaxial growth of the parent compound before the reduction, the Ruddlesden-Popper $R_{n+1}$Ni$_n$O$_{3n+1}$, up to $n=5$ ($R=$La~\cite{Z_Li_2020} and $R$=Nd~\cite{W_Sun_2021}).

\subsection{$d^9$ nickelates without self-doping}

One of the major differences between the nickelates and cuprates is the presence of self-doping in the former. 
If the self-doping is eliminated, the Fermi surface topology of the nickelates will become more similar to the cuprates.

Ref.~\cite{Hirayama_2020} systematically proposed layered nickelates by changing the composition of the layers between the NiO$_2$ layers (``block layer''~\cite{Tokura_1990}, in the case of $R$NiO$_2$, the rare-earth layer corresponds to the block layer).
Ref.~\cite{Pardo_arXiv} also introduced a similar idea:  Indeed, one of the materials studied in Ref.~\cite{Pardo_arXiv} is also investigated in Ref.~\cite{Hirayama_2020}.
In order to avoid charge transfer between the block and NiO$_2$ layers, the material design was carried out using elements in the 1-3 groups that prefer closed shells. 
Furthermore, to keep a large crystal field splitting in the nickel $3d$ manifold, block layers without apical oxygens were considered. 
The structural stability of the designed nickelates was investigated from first-principles~\cite{Hirayama_2020}.

The designed nickelates tend to suppress the self-doping. 
Some of them are completely free from the self-doping, and only the nickel $3d_{x^2-y^2}$ band crosses the Fermi level in the DFT band structure~\cite{Hirayama_2020}. 
The crystal and band structures of two such materials are shown in Fig.~\ref{Fig_d9_nickel}.
In contrast with $R$NiO$_2$, these nickelates may satisfy all the keywords for cuprate analogs listed in Sec.~\ref{sec_intro}, ``two-dimensional,'' ``square lattice,'' ``single orbital,'' and ``near half-filling''.


Since the strongly-correlated nickel $3d_{x^2-y^2}$ orbital becomes half-filling, the proposed self-doping-free $d^9$ nickelates are expected to become a Mott insulator, unlike the infinite-layer nickelates, which could not become a Mott insulator due to the self-doping~\cite{Nomura_2020}.
The effective model for a Mott insulator is the spin-1/2 Heisenberg model with magnetic exchange interaction $J$.  
A first-principles estimate of the $J$ value of the $d^9$ nickelates gives $J=80$-100 meV~\cite{Nomura_2020}, which is not far smaller than that of the cuprates ($\sim$ 130 meV~\cite{Lee_Nagaosa_Wen_2006}). 
The new nickelates showing a large magnetic exchange coupling $J$ would provide an interesting playground to search for new superconductors close to the Mott-Hubbard regime in Zaanen-Sawatzky-Allen diagram~\cite{Nomura_2020}. 

In addition, the $d^9$ nickelates may provide a rare example of realizing a two-dimensional square-lattice Hubbard model, if the holes are doped into the $3d_{x^2-y^2}$ orbital~\cite{Nomura_2020}. 
Since the phase diagram of the doped Hubbard model on the square lattice is being reinvestigated thanks to a recent rapid advance in numerical techniques~\cite{Zheng_2017,Darmawan_2018,HC_Jiang_2019}, 
the $d^9$ nickelates are of great interest as ``testbed'' materials~\cite{Nomura_2020}. 

We note a remarkable experimental advance also in synthesizing $d^9$ nickelates as well as multi-layer square-planar nickelates. 
Independently from the theoretical proposal~\cite{Hirayama_2020}, an experimental work~\cite{Wissel_2020} published around the same time has reported a successful synthesis of T'-type structure La$_2$NiO$_3$F. 
The same material has been investigated in Ref.~\cite{Hirayama_2020} and is predicted to be dynamically stable; thus, the experiment and theoretical prediction show a nice agreement.   
There is also an experimental report of the synthesis of a $d^8$ nickelate Sr$_2$NiO$_2$Cl$_2$~\cite{Tsujimoto_2014}.
This is of great interest because Sr$_2$NiO$_2$Cl$_2$ has the same T-type structure as a theoretically proposed $d^9$ nickelate $A_2$NiO$_2$Br$_2$ ($A$ is a $2.5+$-valent cation) in Fig.~\ref{Fig_d9_nickel}(c), though the nickel valence is different  (for the discussion of $d^8$ nickelates, see Sec.~\ref{Sec:d8_nickelates}).

Another route for realizing $d^9$ nickelates would be nickel fluorides instead of nickel oxides~\cite{Bernardini_2020b}. 
Ref.~\cite{Bernardini_2020b} investigated infinite-layer fluoro-nickelates $A$NiF$_2$ with $A=$ Li, Na, K, Rb, and Cs. Because fluorine is the most electronegative element, the energy difference between the nickel $3d$ and fluorine $2p$ orbitals becomes large. 
This places the fluoro-nickelates well inside the Mott-Hubbard regime and makes the nickel $3d_{x^2-y^2}$ bandwidth smaller (more strongly correlated).

There is also a proposal for nickel chalcogenides~\cite{ZJ_Lang_2022}. 
Ref.~\cite{ZJ_Lang_2022} argues that, compared to nickelates, nickel chalcogenides, such as NdNiS$_2$, show more similar low-energy physics to cuprates and have stronger magnetic exchange coupling.

\subsection{Other variants}

\subsubsection{$d^8$ nickelates}
\label{Sec:d8_nickelates}

The superconductivity in the infinite-layer nickelates emerges close to nickel $d^9$ filling, and thus the relation with the cuprates has been discussed intensively.
Recently, there was a report on the superconductivity in nominally heavily-overdoped compound Ba$_2$CuO$_{3+\delta}$~\cite{WM_Li_2019}.
Then, it would be interesting to expand the space for materials search. 
Refs.~\cite{HS_Jin_2020,Kitamine_2020} have investigated the layered nickelates around $d^8$ filling and discussed possible superconductivity.

\subsubsection{Palladium and silver compounds}

As is discussed in Sec.~\ref{sec_ele_structure}, the bandwidth of nickel $3d_{x^2-y^2}$ orbital (or more precisely, anti-bonding orbital between nickel $3d_{x^2-y^2}$ orbital and oxygen $2p$ orbitals) is smaller compared to that of the cuprates. 
This makes the kinetic energy scale for the superconductivity smaller. 
To get a larger kinetic energy scale, palladium oxides (palladates), in which palladium $4d$ orbitals are more extended in space than nickel $3d$ orbitals, might be promising. 
Layered palladates around $d^9$ filling have been theoretically investigated in Refs.~\cite{Botana_2018} and \cite{Hirayama_2020}.

Silver-based compounds may also be interesting. 
A difficulty in using silver is that silver cation strongly favors 1+ valence. 
Using elements more electronegative than oxygen, Ag$^{2+}$ can be realized. 
Ref.~\cite{Gawraczynski_2019} studied the property of AgF$_2$ and shown that AgF$_2$ can be a cuprate-analog $d^9$ material with a large magnetic exchange coupling, reaching about 70 \% of that of cuprates.

\section{Summary and Outlook}
\label{sec_summary}

The discovery of superconductivity in the doped infinite-layer nickelate Nd$_{0.8}$Sr$_{0.2}$NiO$_2$ was reported in August 2019.
As a review at the initial stage of the research, we have reviewed the basics of the bulk electronic state. 

In Sec.~\ref{sec_ele_structure}, we have discussed that the infinite-layer nickelate $R$NiO$_2$ is similar to the cuprates in that the nickel $3d_{x^2-y^2}$ orbital forms a strongly correlated system on the two-dimensional square lattice near the half-filling.
However, a crucial difference from the cuprates is the presence of the self-doping: rare-earth-layer orbitals form additional Fermi pockets. 
In this sense, the parent compounds of the infinite-layer nickelates could be viewed as a material that fails to be a Mott insulator because of the self-doping. 
The superconductivity is most likely unconventional, and the correlation effect will play an important role.

Then, a question is what kind of correlation physics is essential (Sec.~\ref{sec_minimal_model}).  
If the itinerant rare-earth-layer orbitals are active, they would give rise to Kondo-like physics. 
If the holes are doped into oxygen $2p$ orbitals, the physics of Zhang-Rice singlet may emerge (Mott-Hubbard vs charge-transfer).
When the holes reside within $3d$ manifold (Mott-Hubbard type), there should be a competition between the crystal-field splitting and Hund's coupling (low-spin vs high-spin). 
If the high-spin state is favored, Hund's physics comes into play. 
On the other hand, if the low-spin state is favored, Mott physics possibly masked by the self-doping would be important.
We need more comprehensive experimental and theoretical investigations to clarify these points.

As for a future perspective, as one of the most important tasks, bulk superconducting samples are highly desired. 
If the superconductivity can only be realized in thin films, a difference between thin film and bulk samples needs to be discussed more seriously.

Another crucial question to be elucidated is the possibility of increasing $T_{\rm c}$ (currently, maximum $T_{\rm c}$ is about 15 K) in nickel-based superconductors. 
In this respect, the relation between superconductivity and other possible long-range-order instability needs to be clarified in the nickelates. 
In particular, the relation between superconductivity and magnetism would be one of the most urgent tasks to be investigated, considering the fact that the cuprates have an extraordinarily large magnetic exchange coupling of more than 100 meV.
Although experiments detect a signature of magnetic fluctuations, so far, there is no clear evidence for long-range magnetic order in the infinite-layer parent compound $R$NiO$_2$ (see Sec.~\ref{sec_experiment}).
The presence of the self-doping might be the origin of the absence of long-range order (note that the long-range order is quickly suppressed in the hole-doping side of the cuprate phase diagram~\cite{Keimer_2015}). 
In this respect, $d^9$ nickelates without self-doping (see Sec.~\ref{sec_new_materials}), if realized, would be an ideal material to investigate magnetism in the nickelates and compare it with that of the cuprates. 
An expected Mott insulating behavior eliminates the ambiguity in the mapping to spin model, and the magnetic exchange coupling is well defined.

To conclude, for sure, one needs more comprehensive experimental and theoretical efforts to get a complete picture of nickelate superconductivity. 
The nickelates nicely expand the playground for exploring unconventional superconductivity around the cuprate high-$T_{\rm c}$ superconductivity and provide excellent references to be compared with other unconventional superconductors. 
The establishment and understanding of superconductivity in nickelates will shed new light on the unresolved challenge of elucidating the superconducting mechanism in the correlated materials.

\vspace{0.5cm}

\textit{Note added.} 
After submitting the manuscript, we became aware of intriguing experimental progress. 
Recently, superconductivity has been discovered in film samples of a quintuple-layer nickelate Nd$_6$Ni$_5$O$_{12}$~\cite{Pan_2021}. 
This is the first report of nickelate superconductivity other than in infinite-layer compounds (see Sec.~\ref{Sec:multi-layer}).
Also, charge-order and charge-density-wave instabilities in infinite-layer nickelates are discussed in recent works~\cite{Rossi_arXiv,Krieger_arXiv,Tam_arXiv}.
Interestingly, charge stripes, which couple to the spin degrees of freedom, had been found in a trilayer nickelate La$_4$Ni$_3$O$_8$~\cite{J_Zhang_2016,J_Zhang_2019}.
Elucidating the systematic phase diagram of multi-layer and infinite-layer nickelates is a major challenge for the future.

\section*{Acknowledgements}

We are grateful for fruitful discussions with Motoaki Hirayama, Terumasa Tadano, Takuya Nomoto, Motoharu Kitatani, Yoshihide Yoshimoto, and Kazuma Nakamura. 
We acknowledge the financial support by Grant-in-Aids for Scientific Research (JSPS KAKENHI) [Grant No. 20K14423 (YN), 21H01041 (YN), and 19H05825 (RA)] and MEXT as ``Program for Promoting Researches on the Supercomputer Fugaku'' (Basic Science for Emergence and Functionality in Quantum Matter ---Innovative Strongly-Correlated Electron Science by Integration of ``Fugaku'' and Frontier Experiments---) (JPMXP1020200104).

\vspace{1cm}

\bibliographystyle{hunsrt}

\bibliography{main}

\providecommand{\noopsort}[1]{}\providecommand{\singleletter}[1]{#1}%
\begin{thebibliography}{100}

\bibitem{Bednorz_1986}
J.~G. Bednorz and K.~A. M{\"u}ller.
\newblock {Possible high $T_{\rm c}$ superconductivity in the Ba-La-Cu-O
  system}.
\newblock {\em Zeitschrift f{\"u}r Physik B Condensed Matter}, 64(2):189--193,
  1986.

\bibitem{Pickett_1989}
W.E. Pickett, D.~Singh, D.A. Papaconstantopoulos, H.~Krakauer, M.~Cyrot, and
  F.~Cyrot-Lackmann.
\newblock {Theoretical studies of Sr$_2$VO$_4$, a charge conjugate analog of
  La$_2$CuO$_4$}.
\newblock {\em Physica C: Superconductivity and its Applications},
  162-164:1433--1434, 1989.

\bibitem{Imai_2005}
Yoshiki Imai, Igor Solovyev, and Masatoshi Imada.
\newblock {Electronic Structure of Strongly Correlated Systems Emerging from
  Combining Path-Integral Renormalization Group with the Density-Functional
  Approach}.
\newblock {\em Phys. Rev. Lett.}, 95:176405, Oct 2005.

\bibitem{Matsuno_2005}
J.~Matsuno, Y.~Okimoto, M.~Kawasaki, and Y.~Tokura.
\newblock {Variation of the Electronic Structure in Systematically Synthesized
  ${\mathrm{Sr}}_{2}M{\mathrm{O}}_{4}$ ($M=\mathrm{Ti}$, V, Cr, Mn, and Co)}.
\newblock {\em Phys. Rev. Lett.}, 95:176404, Oct 2005.

\bibitem{Weng_2006}
Hongming Weng, Y.~Kawazoe, Xiangang Wan, and Jinming Dong.
\newblock {Electronic structure and optical properties of layered perovskites
  ${\mathrm{Sr}}_{2}M{\mathrm{O}}_{4}$ ($M=\mathrm{Ti}$, V, Cr, and Mn): An ab
  initio study}.
\newblock {\em Phys. Rev. B}, 74:205112, Nov 2006.

\bibitem{Arita_2007}
R.~Arita, A.~Yamasaki, K.~Held, J.~Matsuno, and K.~Kuroki.
\newblock {${\mathrm{Sr}}_{2}\mathrm{V}{\mathrm{O}}_{4}$ and
  ${\mathrm{Ba}}_{2}\mathrm{V}{\mathrm{O}}_{4}$ under pressure: An orbital
  switch and potential ${d}^{1}$ superconductor}.
\newblock {\em Phys. Rev. B}, 75:174521, May 2007.

\bibitem{Chaloupka_2008}
J.~Chaloupka and Giniyat Khaliullin.
\newblock {Orbital Order and Possible Superconductivity in
  ${\mathrm{LaNiO}}_{3}/{\mathrm{LaMO}}_{3}$ Superlattices}.
\newblock {\em Phys. Rev. Lett.}, 100:016404, Jan 2008.

\bibitem{Hansmann_2009}
P.~Hansmann, Xiaoping Yang, A.~Toschi, G.~Khaliullin, O.~K. Andersen, and
  K.~Held.
\newblock {Turning a Nickelate Fermi Surface into a Cupratelike One through
  Heterostructuring}.
\newblock {\em Phys. Rev. Lett.}, 103:016401, Jun 2009.

\bibitem{Han_2011}
M.~J. Han, Xin Wang, C.~A. Marianetti, and A.~J. Millis.
\newblock {Dynamical Mean-Field Theory of Nickelate Superlattices}.
\newblock {\em Phys. Rev. Lett.}, 107:206804, Nov 2011.

\bibitem{Kim_2016}
Y.~K. Kim, N.~H. Sung, J.~D. Denlinger, and B.~J. Kim.
\newblock {Observation of a $d$-wave gap in electron-doped Sr$_2$IrO$_4$}.
\newblock {\em Nature Physics}, 12(1):37--41, 2016.

\bibitem{Yan_2015}
Y.~J. Yan, M.~Q. Ren, H.~C. Xu, B.~P. Xie, R.~Tao, H.~Y. Choi, N.~Lee, Y.~J.
  Choi, T.~Zhang, and D.~L. Feng.
\newblock {Electron-Doped ${\mathrm{Sr}}_{2}{\mathrm{IrO}}_{4}$: An Analogue of
  Hole-Doped Cuprate Superconductors Demonstrated by Scanning Tunneling
  Microscopy}.
\newblock {\em Phys. Rev. X}, 5:041018, Nov 2015.

\bibitem{MArtins_2011}
Cyril Martins, Markus Aichhorn, Lo\"{\i}g Vaugier, and Silke Biermann.
\newblock {Reduced Effective Spin-Orbital Degeneracy and Spin-Orbital Ordering
  in Paramagnetic Transition-Metal Oxides:
  ${\mathrm{Sr}}_{2}{\mathrm{IrO}}_{4}$ versus
  ${\mathrm{Sr}}_{2}{\mathrm{RhO}}_{4}$}.
\newblock {\em Phys. Rev. Lett.}, 107:266404, Dec 2011.

\bibitem{Arita_2012}
R.~Arita, J.~Kune\ifmmode~\check{s}\else \v{s}\fi{}, A.~V. Kozhevnikov, A.~G.
  Eguiluz, and M.~Imada.
\newblock {Ab initio Studies on the Interplay between Spin-Orbit Interaction
  and Coulomb Correlation in ${\mathrm{Sr}}_{2}{\mathrm{IrO}}_{4}$ and
  ${\mathrm{Ba}}_{2}{\mathrm{IrO}}_{4}$}.
\newblock {\em Phys. Rev. Lett.}, 108:086403, Feb 2012.

\bibitem{Zhang_2013}
Hongbin Zhang, Kristjan Haule, and David Vanderbilt.
\newblock {Effective $J\mathbf{=}1/2$ Insulating State in Ruddlesden-Popper
  Iridates: An $\mathrm{LDA}\mathbf{+}\mathrm{DMFT}$ Study}.
\newblock {\em Phys. Rev. Lett.}, 111:246402, Dec 2013.

\bibitem{Li_2019}
Danfeng Li, Kyuho Lee, Bai~Yang Wang, Motoki Osada, Samuel Crossley, Hye~Ryoung
  Lee, Yi~Cui, Yasuyuki Hikita, and Harold~Y. Hwang.
\newblock {Superconductivity in an infinite-layer nickelate}.
\newblock {\em Nature}, 572(7771):624--627, 2019.

\bibitem{Crespin_1983}
Michel Crespin, Pierre Levitz, and Lucien Gatineau.
\newblock {Reduced forms of LaNiO$_3$ perovskite. Part 1.---Evidence for new
  phases: La$_2$Ni$_2$O$_5$ and LaNiO$_2$}.
\newblock {\em J. Chem. Soc.{,} Faraday Trans. 2}, 79:1181--1194, 1983.

\bibitem{Levitz_1983}
Pierre Levitz, Michel Crespin, and Lucien Gatineau.
\newblock {Reduced forms of LaNiO$_3$ perovskite. Part 2.---X-ray structure of
  LaNiO$_2$ and extended X-ray absorption fine structure study: local
  environment of monovalent nickel}.
\newblock {\em J. Chem. Soc.{,} Faraday Trans. 2}, 79:1195--1203, 1983.

\bibitem{Anisimov_1999}
V.~I. Anisimov, D.~Bukhvalov, and T.~M. Rice.
\newblock {Electronic structure of possible nickelate analogs to the cuprates}.
\newblock {\em Phys. Rev. B}, 59:7901--7906, Mar 1999.

\bibitem{Osada_2020}
Motoki Osada, Bai~Yang Wang, Berit~H. Goodge, Kyuho Lee, Hyeok Yoon, Keita
  Sakuma, Danfeng Li, Masashi Miura, Lena~F. Kourkoutis, and Harold~Y. Hwang.
\newblock {A Superconducting Praseodymium Nickelate with Infinite Layer
  Structure}.
\newblock {\em Nano Letters}, 20(8):5735--5740, 08 2020.

\bibitem{Osada_2020_2}
Motoki Osada, Bai~Yang Wang, Kyuho Lee, Danfeng Li, and Harold~Y. Hwang.
\newblock {Phase diagram of infinite layer praseodymium nickelate
  ${\mathrm{Pr}}_{1\ensuremath{-}x}{\mathrm{Sr}}_{x}{\mathrm{NiO}}_{2}$ thin
  films}.
\newblock {\em Phys. Rev. Materials}, 4:121801, Dec 2020.

\bibitem{Osada_2021}
Motoki Osada, Bai~Yang Wang, Berit~H. Goodge, Shannon~P. Harvey, Kyuho Lee,
  Danfeng Li, Lena~F. Kourkoutis, and Harold~Y. Hwang.
\newblock {Nickelate Superconductivity without Rare-Earth Magnetism:
  (La,Sr)NiO$_2$}.
\newblock {\em Advanced Materials}, 33(45):2104083, 2021.

\bibitem{SW_Zeng_2022}
Zeng Shengwei, Li~Changjian, Chow~Lin Er, Cao Yu, Zhang Zhaoting, Tang~Chi Sin,
  Yin Xinmao, Lim~Zhi Shiuh, Hu~Junxiong, Yang Ping, and Ariando Ariando.
\newblock {Superconductivity in infinite-layer nickelate
  La$_{1-x}$Ca$_x$NiO$_2$ thin films}.
\newblock {\em Science Advances}, 8(7):eabl9927, 2022.

\bibitem{Maeno_1994}
Y.~Maeno, H.~Hashimoto, K.~Yoshida, S.~Nishizaki, T.~Fujita, J.~G. Bednorz, and
  F.~Lichtenberg.
\newblock {Superconductivity in a layered perovskite without copper}.
\newblock {\em Nature}, 372(6506):532--534, 1994.

\bibitem{Norman_2020}
M.~R. Norman.
\newblock {Entering the Nickel Age of Superconductivity}.
\newblock {\em Physics}, 13:85, 2020.

\bibitem{Pickett_2021}
Warren~E. Pickett.
\newblock {The dawn of the nickel age of superconductivity}.
\newblock {\em Nature Reviews Physics}, 3(1):7--8, 2021.

\bibitem{J_Zhang_2021}
Junjie Zhang and Xutang Tao.
\newblock {Review on quasi-2D square planar nickelates}.
\newblock {\em CrystEngComm}, 23:3249--3264, 2021.

\bibitem{Botana_2021}
A.~S. Botana, F.~Bernardini, and A.~Cano.
\newblock {Nickelate Superconductors: An Ongoing Dialog between Theory and
  Experiments}.
\newblock {\em Journal of Experimental and Theoretical Physics},
  132(4):618--627, 2021.

\bibitem{Azuma_1992}
M.~Azuma, Z.~Hiroi, M.~Takano, Y.~Bando, and Y.~Takeda.
\newblock {Superconductivity at 110 K in the infinite-layer compound
  (Sr$_{1-x}$Ca$_x$)$_{1-y}$CuO$_2$}.
\newblock {\em Nature}, 356(6372):775--776, 1992.

\bibitem{Momma_2011}
Koichi Momma and Fujio Izumi.
\newblock {{\it VESTA3} for three-dimensional visualization of crystal,
  volumetric and morphology data}.
\newblock {\em J. Appl. Crystallogr.}, 44(6):1272--1276, Dec 2011.

\bibitem{D_Li_2020}
Danfeng Li, Bai~Yang Wang, Kyuho Lee, Shannon~P. Harvey, Motoki Osada, Berit~H.
  Goodge, Lena~F. Kourkoutis, and Harold~Y. Hwang.
\newblock {Superconducting Dome in
  ${\mathrm{Nd}}_{1\ensuremath{-}x}{\mathrm{Sr}}_{x}{\mathrm{NiO}}_{2}$
  Infinite Layer Films}.
\newblock {\em Phys. Rev. Lett.}, 125:027001, Jul 2020.

\bibitem{Zeng_2020}
Shengwei Zeng, Chi~Sin Tang, Xinmao Yin, Changjian Li, Mengsha Li, Zhen Huang,
  Junxiong Hu, Wei Liu, Ganesh~Ji Omar, Hariom Jani, Zhi~Shiuh Lim, Kun Han,
  Dongyang Wan, Ping Yang, Stephen~John Pennycook, Andrew T.~S. Wee, and
  Ariando Ariando.
\newblock {Phase Diagram and Superconducting Dome of Infinite-Layer
  ${\mathrm{Nd}}_{1\ensuremath{-}x}{\mathrm{Sr}}_{x}{\mathrm{NiO}}_{2}$ Thin
  Films}.
\newblock {\em Phys. Rev. Lett.}, 125:147003, Oct 2020.

\bibitem{Lee_2020}
Kyuho Lee, Berit~H. Goodge, Danfeng Li, Motoki Osada, Bai~Yang Wang, Yi~Cui,
  Lena~F. Kourkoutis, and Harold~Y. Hwang.
\newblock {Aspects of the synthesis of thin film superconducting infinite-layer
  nickelates}.
\newblock {\em APL Materials}, 8(4):041107, 2020.

\bibitem{Q_Gu_2020}
Qiangqiang Gu, Yueying Li, Siyuan Wan, Huazhou Li, Wei Guo, Huan Yang, Qing Li,
  Xiyu Zhu, Xiaoqing Pan, Yuefeng Nie, and Hai-Hu Wen.
\newblock {Single particle tunneling spectrum of superconducting
  Nd$_{1-x}$Sr$_x$NiO$_2$ thin films}.
\newblock {\em Nat. Commun.}, 11(1):6027, 2020.

\bibitem{Q_Gao_2021}
Qiang Gao, Yuchen Zhao, Xing-Jiang Zhou, and Zhihai Zhu.
\newblock {Preparation of Superconducting Thin Films of Infinite-Layer
  Nickelate Nd$_{0.8}$Sr$_{0.2}$NiO$_{2}$}.
\newblock {\em Chinese Physics Letters}, 38(7):077401, 2021.

\bibitem{XR_Zhou_2021}
Xiao-Rong Zhou, Ze-Xin Feng, Pei-Xin Qin, Han Yan, Xiao-Ning Wang, Pan Nie,
  Hao-Jiang Wu, Xin Zhang, Hong-Yu Chen, Zi-Ang Meng, Zeng-Wei Zhu, and Zhi-Qi
  Liu.
\newblock {Negligible oxygen vacancies, low critical current density,
  electric-field modulation, in-plane anisotropic and high-field transport of a
  superconducting Nd$_{0.8}$Sr$_{0.2}$NiO$_2$/SrTiO$_3$ heterostructure}.
\newblock {\em Rare Metals}, 2021.

\bibitem{Y_Li_2021}
Yueying Li, Wenjie Sun, Jiangfeng Yang, Xiangbin Cai, Wei Guo, Zhengbin Gu,
  Ye~Zhu, and Yuefeng Nie.
\newblock {Impact of Cation Stoichiometry on the Crystalline Structure and
  Superconductivity in Nickelates}.
\newblock {\em Frontiers in Physics}, 9:443, 2021.

\bibitem{Zhou_2020}
Xiao-Rong Zhou, Ze-Xin Feng, Pei-Xin Qin, Han Yan, Shuai Hu, Hui-Xin Guo,
  Xiao-Ning Wang, Hao-Jiang Wu, Xin Zhang, Hong-Yu Chen, Xue-Peng Qiu, and
  Zhi-Qi Liu.
\newblock {Absence of superconductivity in Nd$_{0.8}$Sr$_{0.2}$NiO$_x$ thin
  films without chemical reduction}.
\newblock {\em Rare Metals}, 39(4):368--374, 2020.

\bibitem{Liang_Si_2020}
Liang Si, Wen Xiao, Josef Kaufmann, Jan~M. Tomczak, Yi~Lu, Zhicheng Zhong, and
  Karsten Held.
\newblock {Topotactic Hydrogen in Nickelate Superconductors and Akin
  Infinite-Layer Oxides $AB{\mathrm{O}}_{2}$}.
\newblock {\em Phys. Rev. Lett.}, 124:166402, Apr 2020.

\bibitem{Q_Li_2020}
Qing Li, Chengping He, Jin Si, Xiyu Zhu, Yue Zhang, and Hai-Hu Wen.
\newblock {Absence of superconductivity in bulk Nd$_{1-x}$Sr$_x$NiO$_2$}.
\newblock {\em Communications Materials}, 1(1):16, 2020.

\bibitem{BX_Wang_2020}
Bi-Xia Wang, Hong Zheng, E.~Krivyakina, O.~Chmaissem, Pietro~Papa Lopes, J.~W.
  Lynn, Leighanne~C. Gallington, Y.~Ren, S.~Rosenkranz, J.~F. Mitchell, and
  D.~Phelan.
\newblock {Synthesis and characterization of bulk
  ${\mathrm{Nd}}_{1\ensuremath{-}x}{\mathrm{Sr}}_{x}\mathrm{Ni}{\mathrm{O}}_{2}$
  and
  ${\mathrm{Nd}}_{1\ensuremath{-}x}{\mathrm{Sr}}_{x}\mathrm{Ni}{\mathrm{O}}_{3}$}.
\newblock {\em Phys. Rev. Materials}, 4:084409, Aug 2020.

\bibitem{C_He_2021}
Chengping He, Xue Ming, Qing Li, Xiyu Zhu, Jin Si, and Hai-Hu Wen.
\newblock {Synthesis and physical properties of perovskite
  Sm$_{1-x}$Sr$_x$NiO$_3$ ($x$ = 0, 0.2) and infinite-layer
  Sm$_{0.8}$Sr$_{0.2}$NiO$_2$ nickelates}.
\newblock {\em J. Phys.: Condens. Matter}, 33(26):265701, may 2021.

\bibitem{Puphal_2021}
Pascal Puphal, Yu-Mi Wu, Katrin F{\"u}rsich, Hangoo Lee, Mohammad Pakdaman, Jan
  A.~N. Bruin, J{\"u}rgen Nuss, Y.~Eren Suyolcu, Peter~A. van Aken, Bernhard
  Keimer, Masahiko Isobe, and Matthias Hepting.
\newblock {Topotactic transformation of single crystals: From perovskite to
  infinite-layer nickelates}.
\newblock {\em Sci. Adv.}, 7(49):eabl8091, 2021.

\bibitem{Malyi_2022}
Oleksandr~I. Malyi, Julien Varignon, and Alex Zunger.
\newblock {Bulk $\mathrm{NdNi}{\mathrm{O}}_{2}$ is thermodynamically unstable
  with respect to decomposition while hydrogenation reduces the instability and
  transforms it from metal to insulator}.
\newblock {\em Phys. Rev. B}, 105:014106, Jan 2022.

\bibitem{Hsu_2021}
Yu-Te Hsu, Bai~Yang Wang, Maarten Berben, Danfeng Li, Kyuho Lee, Caitlin Duffy,
  Thom Ottenbros, Woo~Jin Kim, Motoki Osada, Steffen Wiedmann, Harold~Y. Hwang,
  and Nigel~E. Hussey.
\newblock {Insulator-to-metal crossover near the edge of the superconducting
  dome in
  ${\mathrm{Nd}}_{1\ensuremath{-}x}{\mathrm{Sr}}_{x}{\mathrm{NiO}}_{2}$}.
\newblock {\em Phys. Rev. Research}, 3:L042015, Nov 2021.

\bibitem{Hayward_2003}
M.A. Hayward and M.J. Rosseinsky.
\newblock {Synthesis of the infinite layer Ni(I) phase NdNiO$_{2+x}$ by low
  temperature reduction of NdNiO$_{3}$ with sodium hydride}.
\newblock {\em Solid State Sci.}, 5(6):839 -- 850, 2003.
\newblock International Conference on Inorganic Materials 2002.

\bibitem{Zaanen_1985}
J.~Zaanen, G.~A. Sawatzky, and J.~W. Allen.
\newblock {Band gaps and electronic structure of transition-metal compounds}.
\newblock {\em Phys. Rev. Lett.}, 55:418--421, Jul 1985.

\bibitem{Hepting_2020}
M.~Hepting, D.~Li, C.~J. Jia, H.~Lu, E.~Paris, Y.~Tseng, X.~Feng, M.~Osada,
  E.~Been, Y.~Hikita, Y.~D. Chuang, Z.~Hussain, K.~J. Zhou, A.~Nag,
  M.~Garcia-Fernandez, M.~Rossi, H.~Y. Huang, D.~J. Huang, Z.~X. Shen,
  T.~Schmitt, H.~Y. Hwang, B.~Moritz, J.~Zaanen, T.~P. Devereaux, and W.~S.
  Lee.
\newblock {Electronic structure of the parent compound of superconducting
  infinite-layer nickelates}.
\newblock {\em Nature Materials}, 19(4):381--385, 2020.

\bibitem{Fu_arXiv}
Ying Fu, Le~Wang, Hu~Cheng, Shenghai Pei, Xuefeng Zhou, Jian Chen, Shaoheng
  Wang, Ran Zhao, Wenrui Jiang, Cai Liu, Mingyuan Huang, XinWei Wang, Yusheng
  Zhao, Dapeng Yu, Fei Ye, Shanmin Wang, and Jia-Wei Mei.
\newblock {Core-level x-ray photoemission and Raman spectroscopy studies on
  electronic structures in Mott-Hubbard type nickelate oxide NdNiO$_2$}, 2019,
  arXiv:1911.03177.

\bibitem{Goodge_2021}
Berit~H. Goodge, Danfeng Li, Kyuho Lee, Motoki Osada, Bai~Yang Wang, George~A.
  Sawatzky, Harold~Y. Hwang, and Lena~F. Kourkoutis.
\newblock {Doping evolution of the Mott{\textendash}Hubbard landscape in
  infinite-layer nickelates}.
\newblock {\em Proceedings of the National Academy of Sciences},
  118(2):e2007683118, 2021.

\bibitem{Z_Chen_arXiv}
Zhuoyu Chen, Motoki Osada, Danfeng Li, Emily~M. Been, Su-Di Chen, Makoto
  Hashimoto, Donghui Lu, Sung-Kwan Mo, Kyuho Lee, Bai~Yang Wang, Fanny
  Rodolakis, Jessica~L. McChesney, Chunjing Jia, Brian Moritz, Thomas~P.
  Devereaux, Harold~Y. Hwang, and Zhi-Xun Shen.
\newblock {Electronic structure of superconducting nickelates probed by
  resonant photoemission spectroscopy}, 2021, arXiv:2106.03963.

\bibitem{H_Lin_2021}
Hai Lin, Dariusz~Jakub Gawryluk, Yannick~Maximilian Klein, Shangxiong Huangfu,
  Ekaterina Pomjakushina, Fabian von Rohr, and Andreas Schilling.
\newblock {Universal spin-glass behaviour in bulk LaNiO$_2$, PrNiO$_2$ and
  NdNiO$_2$}.
\newblock {\em New Journal of Physics}, 2021.

\bibitem{Y_Cui_2021}
Yi~Cui, Cong Li, Qing Li, Xiyu Zhu, Ze~Hu, Yi~feng Yang, Jinshan Zhang, Rong
  Yu, Hai-Hu Wen, and Weiqiang Yu.
\newblock {NMR Evidence of Antiferromagnetic Spin Fluctuations in
  Nd$_{0.85}$Sr$_{0.15}$NiO$_2$}.
\newblock {\em Chinese Physics Letters}, 38(6):067401, 2021.

\bibitem{D_Zhao_2021}
D.~Zhao, Y.~B. Zhou, Y.~Fu, L.~Wang, X.~F. Zhou, H.~Cheng, J.~Li, D.~W. Song,
  S.~J. Li, B.~L. Kang, L.~X. Zheng, L.~P. Nie, Z.~M. Wu, M.~Shan, F.~H. Yu,
  J.~J. Ying, S.~M. Wang, J.~W. Mei, T.~Wu, and X.~H. Chen.
\newblock {Intrinsic Spin Susceptibility and Pseudogaplike Behavior in
  Infinite-Layer ${\mathrm{LaNiO}}_{2}$}.
\newblock {\em Phys. Rev. Lett.}, 126:197001, May 2021.

\bibitem{Lu_2021}
H.~Lu, M.~Rossi, A.~Nag, M.~Osada, D.~F. Li, K.~Lee, B.~Y. Wang,
  M.~Garcia-Fernandez, S.~Agrestini, Z.~X. Shen, E.~M. Been, B.~Moritz, T.~P.
  Devereaux, J.~Zaanen, H.~Y. Hwang, Ke-Jin Zhou, and W.~S. Lee.
\newblock {Magnetic excitations in infinite-layer nickelates}.
\newblock {\em Science}, 373(6551):213--216, 2021.

\bibitem{Adhikary_2020}
Priyo Adhikary, Subhadeep Bandyopadhyay, Tanmoy Das, Indra Dasgupta, and
  Tanusri Saha-Dasgupta.
\newblock {Orbital-selective superconductivity in a two-band model of
  infinite-layer nickelates}.
\newblock {\em Phys. Rev. B}, 102:100501, Sep 2020.

\bibitem{Z_Wang_2020}
Zhan Wang, Guang-Ming Zhang, Yi-feng Yang, and Fu-Chun Zhang.
\newblock {Distinct pairing symmetries of superconductivity in infinite-layer
  nickelates}.
\newblock {\em Phys. Rev. B}, 102:220501, Dec 2020.

\bibitem{Kitamine_2020}
Naoya Kitamine, Masayuki Ochi, and Kazuhiko Kuroki.
\newblock {Designing nickelate superconductors with ${d}^{8}$ configuration
  exploiting mixed-anion strategy}.
\newblock {\em Phys. Rev. Research}, 2:042032, Nov 2020.

\bibitem{X_Wu_arXiv}
Xianxin Wu, Kun Jiang, Domenico~Di Sante, Werner Hanke, A.~P. Schnyder,
  Jiangping Hu, and Ronny Thomale.
\newblock {Surface $s$-wave superconductivity for oxide-terminated
  infinite-layer nickelates}, 2020, arXiv:2008.06009.

\bibitem{BY_Wang_2021}
Bai~Yang Wang, Danfeng Li, Berit~H. Goodge, Kyuho Lee, Motoki Osada, Shannon~P.
  Harvey, Lena~F. Kourkoutis, Malcolm~R. Beasley, and Harold~Y. Hwang.
\newblock {Isotropic Pauli-limited superconductivity in the infinite-layer
  nickelate Nd$_{0.775}$Sr$_{0.225}$NiO$_2$}.
\newblock {\em Nat. Phys.}, 17(4):473--477, 2021.

\bibitem{Y_Xiang_2021}
Ying Xiang, Qing Li, Yueying Li, Huan Yang, Yuefeng Nie, and Hai-Hu Wen.
\newblock {Physical Properties Revealed by Transport Measurements for
  Superconducting Nd$_{0.8}$Sr$_{0.2}$NiO$_2$ Thin Films}.
\newblock {\em Chinese Phys. Lett.}, 38(4):047401, may 2021.

\bibitem{Rossi_2021}
M.~Rossi, H.~Lu, A.~Nag, D.~Li, M.~Osada, K.~Lee, B.~Y. Wang, S.~Agrestini,
  M.~Garcia-Fernandez, J.~J. Kas, Y.-D. Chuang, Z.~X. Shen, H.~Y. Hwang,
  B.~Moritz, Ke-Jin Zhou, T.~P. Devereaux, and W.~S. Lee.
\newblock {Orbital and spin character of doped carriers in infinite-layer
  nickelates}.
\newblock {\em Phys. Rev. B}, 104:L220505, Dec 2021.

\bibitem{SW_Zeng_2022_2}
S.~W. Zeng, X.~M. Yin, C.~J. Li, L.~E. Chow, C.~S. Tang, K.~Han, Z.~Huang,
  Y.~Cao, D.~Y. Wan, Z.~T. Zhang, Z.~S. Lim, C.~Z. Diao, P.~Yang, A.~T.~S. Wee,
  S.~J. Pennycook, and A.~Ariando.
\newblock {Observation of perfect diamagnetism and interfacial effect on the
  electronic structures in infinite layer Nd0.8Sr0.2NiO2 superconductors}.
\newblock {\em Nature Communications}, 13(1):743, 2022.

\bibitem{Nomura_2019}
Yusuke Nomura, Motoaki Hirayama, Terumasa Tadano, Yoshihide Yoshimoto, Kazuma
  Nakamura, and Ryotaro Arita.
\newblock {Formation of a two-dimensional single-component correlated electron
  system and band engineering in the nickelate superconductor {NdNiO$_{2}$}}.
\newblock {\em Phys. Rev. B}, 100:205138, Nov 2019.

\bibitem{Kawamura_2019}
Mitsuaki Kawamura.
\newblock {FermiSurfer: Fermi-surface viewer providing multiple representation
  schemes}.
\newblock {\em Comput. Phys. Commun.}, 239:197 -- 203, 2019.

\bibitem{Lee_2004}
K.-W. Lee and W.~E. Pickett.
\newblock {Infinite-layer $\mathrm{La}\mathrm{Ni}{\mathrm{O}}_{2}$:
  ${\mathrm{Ni}}^{1+}$ is not ${\mathrm{Cu}}^{2+}$}.
\newblock {\em Phys. Rev. B}, 70:165109, Oct 2004.

\bibitem{Botana_2020}
A.~S. Botana and M.~R. Norman.
\newblock {Similarities and Differences between {${\mathrm{LaNiO}}_{2}$ and
  ${\mathrm{CaCuO}}_{2}$} and Implications for Superconductivity}.
\newblock {\em Phys. Rev. X}, 10:011024, Feb 2020.

\bibitem{Sakakibara_2020}
Hirofumi Sakakibara, Hidetomo Usui, Katsuhiro Suzuki, Takao Kotani, Hideo Aoki,
  and Kazuhiko Kuroki.
\newblock {Model Construction and a Possibility of Cupratelike Pairing in a New
  ${d}^{9}$ Nickelate Superconductor
  $(\mathrm{Nd},\mathrm{Sr}){\mathrm{NiO}}_{2}$}.
\newblock {\em Phys. Rev. Lett.}, 125:077003, Aug 2020.

\bibitem{Wu_2020}
Xianxin Wu, Domenico Di~Sante, Tilman Schwemmer, Werner Hanke, Harold~Y. Hwang,
  Srinivas Raghu, and Ronny Thomale.
\newblock {Robust ${d}_{{x}^{2}\ensuremath{-}{y}^{2}}$-wave superconductivity
  of infinite-layer nickelates}.
\newblock {\em Phys. Rev. B}, 101:060504, Feb 2020.

\bibitem{Bernardini_2020c}
Fabio Bernardini and Andres Cano.
\newblock {Stability and electronic properties of LaNiO$_2$/SrTiO$_3$
  heterostructures}.
\newblock {\em Journal of Physics: Materials}, 2020.

\bibitem{Geisler_2020}
Benjamin Geisler and Rossitza Pentcheva.
\newblock {Fundamental difference in the electronic reconstruction of
  infinite-layer versus perovskite neodymium nickelate films on
  ${\mathrm{SrTiO}}_{3}$(001)}.
\newblock {\em Phys. Rev. B}, 102:020502, Jul 2020.

\bibitem{Geisler_2021}
Benjamin Geisler and Rossitza Pentcheva.
\newblock {Correlated interface electron gas in infinite-layer nickelate versus
  cuprate films on ${\mathrm{SrTiO}}_{3}(001)$}.
\newblock {\em Phys. Rev. Research}, 3:013261, Mar 2021.

\bibitem{He_2020}
Ri~He, Peiheng Jiang, Yi~Lu, Yidao Song, Mingxing Chen, Mingliang Jin, Lingling
  Shui, and Zhicheng Zhong.
\newblock {Polarity-induced electronic and atomic reconstruction at
  ${\mathrm{NdNiO}}_{2}/{\mathrm{SrTiO}}_{3}$ interfaces}.
\newblock {\em Phys. Rev. B}, 102:035118, Jul 2020.

\bibitem{Y_Zhang_2020}
Yang Zhang, Ling-Fang Lin, Wenjun Hu, Adriana Moreo, Shuai Dong, and Elbio
  Dagotto.
\newblock {Similarities and differences between nickelate and cuprate films
  grown on a ${\mathrm{SrTiO}}_{3}$ substrate}.
\newblock {\em Phys. Rev. B}, 102:195117, Nov 2020.

\bibitem{Siegrist_1988}
T.~Siegrist, S.~M. Zahurak, D.~W. Murphy, and R.~S. Roth.
\newblock {The parent structure of the layered high-temperature
  superconductors}.
\newblock {\em Nature}, 334(6179):231--232, 1988.

\bibitem{Held_2007}
K.~Held.
\newblock {Electronic structure calculations using dynamical mean field
  theory}.
\newblock {\em Advances in Physics}, 56(6):829--926, 2007.

\bibitem{Y_Gu_2020}
Yuhao Gu, Sichen Zhu, Xiaoxuan Wang, Jiangping Hu, and Hanghui Chen.
\newblock {A substantial hybridization between correlated Ni-$d$ orbital and
  itinerant electrons in infinite-layer nickelates}.
\newblock {\em Communications Physics}, 3(1):84, 2020.

\bibitem{Hirayama_2020}
Motoaki Hirayama, Terumasa Tadano, Yusuke Nomura, and Ryotaro Arita.
\newblock {Materials design of dynamically stable ${d}^{9}$ layered
  nickelates}.
\newblock {\em Phys. Rev. B}, 101:075107, Feb 2020.

\bibitem{Been_2021}
Emily Been, Wei-Sheng Lee, Harold~Y. Hwang, Yi~Cui, Jan Zaanen, Thomas
  Devereaux, Brian Moritz, and Chunjing Jia.
\newblock {Electronic Structure Trends Across the Rare-Earth Series in
  Superconducting Infinite-Layer Nickelates}.
\newblock {\em Phys. Rev. X}, 11:011050, Mar 2021.

\bibitem{Kapeghian_2020}
Jesse Kapeghian and Antia~S. Botana.
\newblock {Electronic structure and magnetism in infinite-layer nickelates
  $R{\mathrm{NiO}}_{2}$ ($R=\mathrm{La}$-$\mathrm{Lu}$)}.
\newblock {\em Phys. Rev. B}, 102:205130, Nov 2020.

\bibitem{P_Jiang_2019}
Peiheng Jiang, Liang Si, Zhaoliang Liao, and Zhicheng Zhong.
\newblock {Electronic structure of rare-earth infinite-layer
  $R\mathrm{Ni}{\mathrm{O}}_{2}(R=\mathrm{La},\mathrm{Nd})$}.
\newblock {\em Phys. Rev. B}, 100:201106, Nov 2019.

\bibitem{Choi_2020}
Mi-Young Choi, Kwan-Woo Lee, and Warren~E. Pickett.
\newblock {Role of $4f$ states in infinite-layer ${\mathrm{NdNiO}}_{2}$}.
\newblock {\em Phys. Rev. B}, 101:020503, Jan 2020.

\bibitem{R_Zhang_2021}
Ruiqi Zhang, Christopher Lane, Bahadur Singh, Johannes Nokelainen, Bernardo
  Barbiellini, Robert~S. Markiewicz, Arun Bansil, and Jianwei Sun.
\newblock {Magnetic and $f$-electron effects in LaNiO$_2$ and NdNiO$_2$
  nickelates with cuprate-like $3d_{x^2-y^2}$ band}.
\newblock {\em Communications Physics}, 4(1):118, 2021.

\bibitem{Bandyopadhyay_2020}
Subhadeep Bandyopadhyay, Priyo Adhikary, Tanmoy Das, Indra Dasgupta, and
  Tanusri Saha-Dasgupta.
\newblock {Superconductivity in infinite-layer nickelates: Role of $f$
  orbitals}.
\newblock {\em Phys. Rev. B}, 102:220502, Dec 2020.

\bibitem{Olevano_2020}
Valerio Olevano, Fabio Bernardini, Xavier Blase, and Andr\'es Cano.
\newblock {Ab initio many-body $GW$ correlations in the electronic structure of
  ${\mathrm{LaNiO}}_{2}$}.
\newblock {\em Phys. Rev. B}, 101:161102, Apr 2020.

\bibitem{Kutepov_2021}
Andrey~L. Kutepov.
\newblock {Electronic structure of $\mathrm{La}\mathrm{Ni}{\mathrm{O}}_{2}$ and
  $\mathrm{Ca}\mathrm{Cu}{\mathrm{O}}_{2}$ from a self-consistent
  vertex-corrected $GW$ approach}.
\newblock {\em Phys. Rev. B}, 104:085109, Aug 2021.

\bibitem{Georges_1996}
Antoine Georges, Gabriel Kotliar, Werner Krauth, and Marcelo~J. Rozenberg.
\newblock {Dynamical mean-field theory of strongly correlated fermion systems
  and the limit of infinite dimensions}.
\newblock {\em Rev. Mod. Phys.}, 68:13--125, Jan 1996.

\bibitem{Kotliar_2006}
G.~Kotliar, S.~Y. Savrasov, K.~Haule, V.~S. Oudovenko, O.~Parcollet, and C.~A.
  Marianetti.
\newblock {Electronic structure calculations with dynamical mean-field theory}.
\newblock {\em Rev. Mod. Phys.}, 78:865--951, Aug 2006.

\bibitem{Ryee_2020}
Siheon Ryee, Hongkee Yoon, Taek~Jung Kim, Min~Yong Jeong, and Myung~Joon Han.
\newblock {Induced magnetic two-dimensionality by hole doping in the
  superconducting infinite-layer nickelate
  ${\mathrm{Nd}}_{1\ensuremath{-}x}{\mathrm{Sr}}_{x}{\mathrm{NiO}}_{2}$}.
\newblock {\em Phys. Rev. B}, 101:064513, Feb 2020.

\bibitem{Werner_2020}
Philipp Werner and Shintaro Hoshino.
\newblock {Nickelate superconductors: Multiorbital nature and spin freezing}.
\newblock {\em Phys. Rev. B}, 101:041104, Jan 2020.

\bibitem{Kitatani_2020}
Motoharu Kitatani, Liang Si, Oleg Janson, Ryotaro Arita, Zhicheng Zhong, and
  Karsten Held.
\newblock {Nickelate superconductors---a renaissance of the one-band Hubbard
  model}.
\newblock {\em npj Quantum Materials}, 5(1):59, 2020.

\bibitem{Lechermann_2020}
Frank Lechermann.
\newblock {Late transition metal oxides with infinite-layer structure:
  Nickelates versus cuprates}.
\newblock {\em Phys. Rev. B}, 101:081110, Feb 2020.

\bibitem{Lechermann_2020b}
Frank Lechermann.
\newblock {Multiorbital Processes Rule the
  ${\mathrm{Nd}}_{1\ensuremath{-}x}{\mathrm{Sr}}_{x}{\mathrm{NiO}}_{2}$ Normal
  State}.
\newblock {\em Phys. Rev. X}, 10:041002, Oct 2020.

\bibitem{Lechermann_2021}
Frank Lechermann.
\newblock {Doping-dependent character and possible magnetic ordering of
  ${\mathrm{NdNiO}}_{2}$}.
\newblock {\em Phys. Rev. Materials}, 5:044803, Apr 2021.

\bibitem{Karp_2020}
Jonathan Karp, Antia~S. Botana, Michael~R. Norman, Hyowon Park, Manuel Zingl,
  and Andrew Millis.
\newblock {Many-Body Electronic Structure of ${\mathrm{NdNiO}}_{2}$ and
  ${\mathrm{CaCuO}}_{2}$}.
\newblock {\em Phys. Rev. X}, 10:021061, Jun 2020.

\bibitem{Karp_2020b}
Jonathan Karp, Alexander Hampel, Manuel Zingl, Antia~S. Botana, Hyowon Park,
  Michael~R. Norman, and Andrew~J. Millis.
\newblock {Comparative many-body study of
  ${\mathrm{Pr}}_{4}{\mathrm{Ni}}_{3}{\mathrm{O}}_{8}$ and
  ${\mathrm{NdNiO}}_{2}$}.
\newblock {\em Phys. Rev. B}, 102:245130, Dec 2020.

\bibitem{Karp_2021}
Jonathan Karp, Alexander Hampel, and Andrew~J. Millis.
\newblock {Dependence of $\mathrm{DFT}+\mathrm{DMFT}$ results on the
  construction of the correlated orbitals}.
\newblock {\em Phys. Rev. B}, 103:195101, May 2021.

\bibitem{Leonov_2020}
I.~Leonov, S.~L. Skornyakov, and S.~Y. Savrasov.
\newblock {Lifshitz transition and frustration of magnetic moments in
  infinite-layer ${\mathrm{NdNiO}}_{2}$ upon hole doping}.
\newblock {\em Phys. Rev. B}, 101:241108, Jun 2020.

\bibitem{Leonov_2021}
I.~Leonov.
\newblock {Effect of lattice strain on the electronic structure and magnetic
  correlations in infinite-layer (Nd,Sr)NiO$_2$}.
\newblock {\em Journal of Alloys and Compounds}, 883:160888, 2021.

\bibitem{X_Wan_2021}
Xiangang Wan, Vsevolod Ivanov, Giacomo Resta, Ivan Leonov, and Sergey~Y.
  Savrasov.
\newblock {Exchange interactions and sensitivity of the Ni two-hole spin state
  to Hund's coupling in doped ${\mathrm{NdNiO}}_{2}$}.
\newblock {\em Phys. Rev. B}, 103:075123, Feb 2021.

\bibitem{Y_Wang_2020}
Y.~Wang, C.-J. Kang, H.~Miao, and G.~Kotliar.
\newblock {Hund's metal physics: From ${\mathrm{SrNiO}}_{2}$ to
  ${\mathrm{LaNiO}}_{2}$}.
\newblock {\em Phys. Rev. B}, 102:161118, Oct 2020.

\bibitem{Kang_2021}
Chang-Jong Kang and Gabriel Kotliar.
\newblock {Optical Properties of the Infinite-Layer
  ${\mathrm{La}}_{1\ensuremath{-}x}{\mathrm{Sr}}_{x}{\mathrm{NiO}}_{2}$ and
  Hidden Hund's Physics}.
\newblock {\em Phys. Rev. Lett.}, 126:127401, Mar 2021.

\bibitem{Z_Liu_2021}
Zhao Liu, Chenchao Xu, Chao Cao, W.~Zhu, Z.~F. Wang, and Jinlong Yang.
\newblock {Doping dependence of electronic structure of infinite-layer
  ${\mathrm{NdNiO}}_{2}$}.
\newblock {\em Phys. Rev. B}, 103:045103, Jan 2021.

\bibitem{Higashi_2021}
Keisuke Higashi, Mathias Winder, Jan Kune\ifmmode~\check{s}\else \v{s}\fi{},
  and Atsushi Hariki.
\newblock {Core-Level X-Ray Spectroscopy of Infinite-Layer Nickelate:
  $\mathrm{LDA}+\mathrm{DMFT}$ Study}.
\newblock {\em Phys. Rev. X}, 11:041009, Oct 2021.

\bibitem{Petocchi_2020}
Francesco Petocchi, Viktor Christiansson, Fredrik Nilsson, Ferdi Aryasetiawan,
  and Philipp Werner.
\newblock {Normal State of
  ${\mathrm{Nd}}_{1\ensuremath{-}x}{\mathrm{Sr}}_{x}{\mathrm{NiO}}_{2}$ from
  Self-Consistent $GW+\mathrm{EDMFT}$}.
\newblock {\em Phys. Rev. X}, 10:041047, Dec 2020.

\bibitem{B_Kang_arxiv}
Byungkyun Kang, Corey Melnick, Patrick Semon, Siheon Ryee, Myung~Joon Han,
  Gabriel Kotliar, and Sangkook Choi.
\newblock {Infinite-layer nickelates as Ni-$e_g$ Hund's metals}, 2021,
  arXiv:2007.14610.

\bibitem{Hayward_1999}
M.~A. Hayward, M.~A. Green, M.~J. Rosseinsky, and J.~Sloan.
\newblock {Sodium Hydride as a Powerful Reducing Agent for Topotactic Oxide
  Deintercalation: Synthesis and Characterization of the Nickel(I) Oxide
  LaNiO$_{2}$}.
\newblock {\em J. Am. Chem. Soc.}, 121(38):8843--8854, 09 1999.

\bibitem{HuZhang_2020}
Hu~Zhang, Lipeng Jin, Shanmin Wang, Bin Xi, Xingqiang Shi, Fei Ye, and Jia-Wei
  Mei.
\newblock {Effective Hamiltonian for nickelate oxides
  ${\mathrm{Nd}}_{1\ensuremath{-}x}{\mathrm{Sr}}_{x}{\mathrm{NiO}}_{2}$}.
\newblock {\em Phys. Rev. Research}, 2:013214, Feb 2020.

\bibitem{Z_Liu_2020}
Zhao Liu, Zhi Ren, Wei Zhu, Zhengfei Wang, and Jinlong Yang.
\newblock {Electronic and magnetic structure of infinite-layer NdNiO$_2$: trace
  of antiferromagnetic metal}.
\newblock {\em npj Quantum Materials}, 5(1):31, 2020.

\bibitem{ZJ_Lang_2021}
Zi-Jian Lang, Ruoshi Jiang, and Wei Ku.
\newblock {Strongly correlated doped hole carriers in the superconducting
  nickelates: Their location, local many-body state, and low-energy effective
  Hamiltonian}.
\newblock {\em Phys. Rev. B}, 103:L180502, May 2021.

\bibitem{Choi_2020b}
Mi-Young Choi, Warren~E. Pickett, and Kwan-Woo Lee.
\newblock {Fluctuation-frustrated flat band instabilities in
  ${\mathrm{NdNiO}}_{2}$}.
\newblock {\em Phys. Rev. Research}, 2:033445, Sep 2020.

\bibitem{Lee_Nagaosa_Wen_2006}
Patrick~A. Lee, Naoto Nagaosa, and Xiao-Gang Wen.
\newblock {Doping a Mott insulator: Physics of high-temperature
  superconductivity}.
\newblock {\em Rev. Mod. Phys.}, 78:17--85, Jan 2006.

\bibitem{Jiang_2020}
Mi~Jiang, Mona Berciu, and George~A. Sawatzky.
\newblock {Critical Nature of the {Ni} Spin State in Doped
  {${\mathrm{NdNiO}}_{2}$}}.
\newblock {\em Phys. Rev. Lett.}, 124:207004, May 2020.

\bibitem{GM_Zhang_2020}
Guang-Ming Zhang, Yi-Feng Yang, and Fu-Chun Zhang.
\newblock {Self-doped Mott insulator for parent compounds of nickelate
  superconductors}.
\newblock {\em Phys. Rev. B}, 101:020501, Jan 2020.

\bibitem{Nomura_2020}
Yusuke Nomura, Takuya Nomoto, Motoaki Hirayama, and Ryotaro Arita.
\newblock {Magnetic exchange coupling in cuprate-analog ${d}^{9}$ nickelates}.
\newblock {\em Phys. Rev. Research}, 2:043144, Oct 2020.

\bibitem{Katukuri_2020}
Vamshi~M. Katukuri, Nikolay~A. Bogdanov, Oskar Weser, Jeroen van~den Brink, and
  Ali Alavi.
\newblock {Electronic correlations and magnetic interactions in infinite-layer
  ${\mathrm{NdNiO}}_{2}$}.
\newblock {\em Phys. Rev. B}, 102:241112, Dec 2020.

\bibitem{Bardeen_1957}
J.~Bardeen, L.~N. Cooper, and J.~R. Schrieffer.
\newblock {Theory of Superconductivity}.
\newblock {\em Phys. Rev.}, 108:1175--1204, Dec 1957.

\bibitem{Hirsch_2019}
J.E. Hirsch and F.~Marsiglio.
\newblock Hole superconductivity in infinite-layer nickelates.
\newblock {\em Physica C: Superconductivity and its Applications}, 566:1353534,
  2019.

\bibitem{Sawatzky_2019}
G.~A. Sawatzky.
\newblock {Superconductivity seen in a non-magnetic nickel oxide}.
\newblock {\em Nature News and Views}, 572:592--593, 2019.

\bibitem{Ikeda_2016}
Ai~Ikeda, Yoshiharu Krockenberger, Hiroshi Irie, Michio Naito, and Hideki
  Yamamoto.
\newblock {Direct observation of infinite NiO$_2$ planes in LaNiO$_2$ films}.
\newblock {\em Applied Physics Express}, 9(6):061101, may 2016.

\bibitem{Hewson_1993}
Alexander~Cyril Hewson.
\newblock {\em {The Kondo Problem to Heavy Fermions}}.
\newblock Cambridge Studies in Magnetism. Cambridge University Press, 1993.

\bibitem{Zhang_1988}
F.~C. Zhang and T.~M. Rice.
\newblock {Effective Hamiltonian for the superconducting Cu oxides}.
\newblock {\em Phys. Rev. B}, 37:3759--3761, Mar 1988.

\bibitem{YH_Zhang_2020}
Ya-Hui Zhang and Ashvin Vishwanath.
\newblock {Type-II $t$-$J$ model in superconducting nickelate
  ${\mathrm{Nd}}_{1\ensuremath{-}x}{\mathrm{Sr}}_{x}{\mathrm{NiO}}_{2}$}.
\newblock {\em Phys. Rev. Research}, 2:023112, May 2020.

\bibitem{LH_Hu_2019}
Lun-Hui Hu and Congjun Wu.
\newblock {Two-band model for magnetism and superconductivity in nickelates}.
\newblock {\em Phys. Rev. Research}, 1:032046, Dec 2019.

\bibitem{J_Chang_2020}
Jun Chang, Jize Zhao, and Yang Ding.
\newblock {Hund-Heisenberg model in superconducting infinite-layer nickelates}.
\newblock {\em The European Physical Journal B}, 93(12):220, 2020.

\bibitem{Matsumoto_2019}
Yuki Matsumoto, Takafumi Yamamoto, Kousuke Nakano, Hiroshi Takatsu, Taito
  Murakami, Kenta Hongo, Ryo Maezono, Hiraku Ogino, Dongjoon Song, Craig~M.
  Brown, C{\'e}dric Tassel, and Hiroshi Kageyama.
\newblock {High-Pressure Synthesis of $A_2$NiO$_2$Ag$_2$Se$_2$ ($A$=Sr, Ba)
  with a High-Spin Ni$^{2+}$ in Square-Planar Coordination}.
\newblock {\em Angew. Chem. Int. Ed.}, 58(3):756--759, 2019.

\bibitem{Nica_2020}
Emilian~M. Nica, Jyoti Krishna, Rong Yu, Qimiao Si, Antia~S. Botana, and Onur
  Erten.
\newblock {Theoretical investigation of superconductivity in trilayer
  square-planar nickelates}.
\newblock {\em Phys. Rev. B}, 102:020504, Jul 2020.

\bibitem{note_Mukuda_2012}
For a review on multi-layer cuprates, see, e.g., Ref.~\cite{Mukuda_2012}.

\bibitem{J_Zhang_2017}
Junjie Zhang, A.~S. Botana, J.~W. Freeland, D.~Phelan, Hong Zheng, V.~Pardo,
  M.~R. Norman, and J.~F. Mitchell.
\newblock {Large orbital polarization in a metallic square-planar nickelate}.
\newblock {\em Nat. Phys.}, 13(9):864--869, 2017.

\bibitem{JQ_Lin_2021}
J.~Q. Lin, P.~Villar~Arribi, G.~Fabbris, A.~S. Botana, D.~Meyers, H.~Miao,
  Y.~Shen, D.~G. Mazzone, J.~Feng, S.~G. Chiuzb\ifmmode~\u{a}\else
  \u{a}\fi{}ian, A.~Nag, A.~C. Walters, M.~Garc\'{\i}a-Fern\'andez, Ke-Jin
  Zhou, J.~Pelliciari, I.~Jarrige, J.~W. Freeland, Junjie Zhang, J.~F.
  Mitchell, V.~Bisogni, X.~Liu, M.~R. Norman, and M.~P.~M. Dean.
\newblock {Strong Superexchange in a ${d}^{9\ensuremath{-}\ensuremath{\delta}}$
  Nickelate Revealed by Resonant Inelastic X-Ray Scattering}.
\newblock {\em Phys. Rev. Lett.}, 126:087001, Feb 2021.

\bibitem{Z_Li_2020}
Z.~Li, W.~Guo, T.~T. Zhang, J.~H. Song, T.~Y. Gao, Z.~B. Gu, and Y.~F. Nie.
\newblock {Epitaxial growth and electronic structure of Ruddlesden--Popper
  nickelates (La$_{n+1}$Ni$_n$O$_{3n+1}$, $n$ = 1--5)}.
\newblock {\em APL Materials}, 8(9):091112, 2020.

\bibitem{W_Sun_2021}
Wenjie Sun, Yueying Li, Xiangbin Cai, Jiangfeng Yang, Wei Guo, Zhengbin Gu,
  Ye~Zhu, and Yuefeng Nie.
\newblock {Electronic and transport properties in Ruddlesden-Popper neodymium
  nickelates ${\mathrm{Nd}}_{n+1}{\mathrm{Ni}}_{n}{\mathrm{O}}_{3n+1}$
  ($n=1$--5)}.
\newblock {\em Phys. Rev. B}, 104:184518, Nov 2021.

\bibitem{Tokura_1990}
Y.~{Tokura} and T.~{Arima}.
\newblock {New classification method for layered copper oxide compounds and its
  application to design of new high Tc superconductors}.
\newblock {\em Jpn. J. Appl. Phys.}, 29:2388--2402, November 1990.

\bibitem{Pardo_arXiv}
Victor Pardo and Antia~S. Botana.
\newblock {Is there a proximate antiferromagnetic insulating phase in
  infinite-layer nickelates?}, 2020, arXiv:2012.02711.

\bibitem{Zheng_2017}
Bo-Xiao Zheng, Chia-Min Chung, Philippe Corboz, Georg Ehlers, Ming-Pu Qin,
  Reinhard~M. Noack, Hao Shi, Steven~R. White, Shiwei Zhang, and Garnet Kin-Lic
  Chan.
\newblock {Stripe order in the underdoped region of the two-dimensional Hubbard
  model}.
\newblock {\em Science}, 358(6367):1155--1160, 2017.

\bibitem{Darmawan_2018}
Andrew~S. Darmawan, Yusuke Nomura, Youhei Yamaji, and Masatoshi Imada.
\newblock {Stripe and superconducting order competing in the Hubbard model on a
  square lattice studied by a combined variational Monte Carlo and tensor
  network method}.
\newblock {\em Phys. Rev. B}, 98:205132, Nov 2018.

\bibitem{HC_Jiang_2019}
Hong-Chen Jiang and Thomas~P. Devereaux.
\newblock {Superconductivity in the doped Hubbard model and its interplay with
  next-nearest hopping $t'$}.
\newblock {\em Science}, 365(6460):1424--1428, 2019.

\bibitem{Wissel_2020}
Kerstin Wissel, Ali~Muhammad Malik, Sami Vasala, Sergi Plana-Ruiz, Ute Kolb,
  Peter~R. Slater, Ivan da~Silva, Lambert Alff, Jochen Rohrer, and Oliver
  Clemens.
\newblock {Topochemical Reduction of La$_2$NiO$_3$F$_2$: The First Ni-Based
  Ruddlesden--Popper $n$=1 T'-Type Structure and the Impact of Reduction on
  Magnetic Ordering}.
\newblock {\em Chemistry of Materials}, 32(7):3160--3179, 04 2020.

\bibitem{Tsujimoto_2014}
Y.~Tsujimoto, C.~I. Sathish, Y.~Matsushita, K.~Yamaura, and T.~Uchikoshi.
\newblock {New members of layered oxychloride perovskites with square planar
  coordination: Sr$_{2}M$O$_{2}$Cl$_{2}$ ($M$ = Mn{,} Ni) and
  Ba$_{2}$PdO$_{2}$Cl$_{2}$}.
\newblock {\em Chem. Commun.}, 50:5915--5918, 2014.

\bibitem{Bernardini_2020b}
F~Bernardini, V~Olevano, X~Blase, and A~Cano.
\newblock {Infinite-layer fluoro-nickelates as $d^9$ model materials}.
\newblock {\em Journal of Physics: Materials}, 3(3):035003, jun 2020.

\bibitem{ZJ_Lang_2022}
Zi-Jian Lang, Ruoshi Jiang, and Wei Ku.
\newblock {Proposal to improve Ni-based superconductors via enhanced charge
  transfer}.
\newblock {\em Phys. Rev. B}, 105:L100501, Mar 2022.

\bibitem{WM_Li_2019}
W.~M. Li, J.~F. Zhao, L.~P. Cao, Z.~Hu, Q.~Z. Huang, X.~C. Wang, Y.~Liu, G.~Q.
  Zhao, J.~Zhang, Q.~Q. Liu, R.~Z. Yu, Y.~W. Long, H.~Wu, H.~J. Lin, C.~T.
  Chen, Z.~Li, Z.~Z. Gong, Z.~Guguchia, J.~S. Kim, G.~R. Stewart, Y.~J. Uemura,
  S.~Uchida, and C.~Q. Jin.
\newblock {Superconductivity in a unique type of copper oxide}.
\newblock {\em Proc. Natl. Acad. Sci. USA}, 116(25):12156--12160, 2019.

\bibitem{HS_Jin_2020}
Hyo-Sun Jin, Warren~E. Pickett, and Kwan-Woo Lee.
\newblock {Proposed ordering of textured spin singlets in a bulk infinite-layer
  nickelate}.
\newblock {\em Phys. Rev. Research}, 2:033197, Aug 2020.

\bibitem{Botana_2018}
A.~S. Botana and M.~R. Norman.
\newblock {Layered palladates and their relation to nickelates and cuprates}.
\newblock {\em Phys. Rev. Materials}, 2:104803, Oct 2018.

\bibitem{Gawraczynski_2019}
Jakub Gawraczy{\'n}ski, Dominik Kurzyd{\l}owski, Russell~A. Ewings,
  Subrahmanyam Bandaru, Wojciech Gadomski, Zoran Mazej, Giampiero Ruani, Ilaria
  Bergenti, Tomasz Jaro{\'n}, Andrew Ozarowski, Stephen Hill, Piotr~J.
  Leszczy{\'n}ski, Kamil Tok{\'a}r, Mariana Derzsi, Paolo Barone, Krzysztof
  Wohlfeld, Jos{\'e} Lorenzana, and Wojciech Grochala.
\newblock {Silver route to cuprate analogs}.
\newblock {\em Proc. Natl. Acad. Sci. USA}, 116(5):1495--1500, 2019.

\bibitem{Keimer_2015}
B.~Keimer, S.~A. Kivelson, M.~R. Norman, S.~Uchida, and J.~Zaanen.
\newblock {From quantum matter to high-temperature superconductivity in copper
  oxides}.
\newblock {\em Nature}, 518(7538):179--186, 2015.

\bibitem{Pan_2021}
Grace~A. Pan, Dan Ferenc~Segedin, Harrison LaBollita, Qi~Song, Emilian~M. Nica,
  Berit~H. Goodge, Andrew~T. Pierce, Spencer Doyle, Steve Novakov, Denisse
  C{\'o}rdova~Carrizales, Alpha~T. N'Diaye, Padraic Shafer, Hanjong Paik,
  John~T. Heron, Jarad~A. Mason, Amir Yacoby, Lena~F. Kourkoutis, Onur Erten,
  Charles~M. Brooks, Antia~S. Botana, and Julia~A. Mundy.
\newblock {Superconductivity in a quintuple-layer square-planar nickelate}.
\newblock {\em Nature Materials}, 2021.

\bibitem{Rossi_arXiv}
Matteo Rossi, Motoki Osada, Jaewon Choi, Stefano Agrestini, Daniel Jost,
  Yonghun Lee, Haiyu Lu, Bai~Yang Wang, Kyuho Lee, Abhishek Nag, Yi-De Chuang,
  Cheng-Tai Kuo, Sang-Jun Lee, Brian Moritz, Thomas~P. Devereaux, Zhi-Xun Shen,
  Jun-Sik Lee, Ke-Jin Zhou, Harold~Y. Hwang, and Wei-Sheng Lee.
\newblock {A Broken Translational Symmetry State in an Infinite-Layer
  Nickelate}, 2021, 2112.02484.

\bibitem{Krieger_arXiv}
G.~Krieger, L.~Martinelli, S.~Zeng, L.~E. Chow, K.~Kummer, R.~Arpaia,
  M.~Moretti Sala, N.~B. Brookes, A.~Ariando, N.~Viart, M.~Salluzzo,
  G.~Ghiringhelli, and D.~Preziosi.
\newblock {Charge and spin order dichotomy in NdNiO$_2$ driven by SrTiO$_3$
  capping layer}, 2021, 2112.03341.

\bibitem{Tam_arXiv}
Charles~C. Tam, Jaewon Choi, Xiang Ding, Stefano Agrestini, Abhishek Nag, Bing
  Huang, Huiqian Luo, Mirian García-Fernández, Liang Qiao, and Ke-Jin Zhou.
\newblock Charge density waves in infinite-layer ndnio$_2$ nickelates, 2021,
  2112.04440.

\bibitem{J_Zhang_2016}
Junjie Zhang, Yu-Sheng Chen, D.~Phelan, Hong Zheng, M.~R. Norman, and J.~F.
  Mitchell.
\newblock {Stacked charge stripes in the quasi-2D trilayer nickelate La4Ni3O8}.
\newblock {\em Proceedings of the National Academy of Sciences},
  113(32):8945--8950, 2016.

\bibitem{J_Zhang_2019}
Junjie Zhang, D.~M. Pajerowski, A.~S. Botana, Hong Zheng, L.~Harriger,
  J.~Rodriguez-Rivera, J.~P.~C. Ruff, N.~J. Schreiber, B.~Wang, Yu-Sheng Chen,
  W.~C. Chen, M.~R. Norman, S.~Rosenkranz, J.~F. Mitchell, and D.~Phelan.
\newblock {Spin Stripe Order in a Square Planar Trilayer Nickelate}.
\newblock {\em Phys. Rev. Lett.}, 122:247201, Jun 2019.

\bibitem{Mukuda_2012}
Hidekazu Mukuda, Sunao Shimizu, Akira Iyo, and Yoshio Kitaoka.
\newblock {High-$T_{\rm c}$ Superconductivity and Antiferromagnetism in
  Multilayered Copper Oxides --A New Paradigm of Superconducting Mechanism--}.
\newblock {\em Journal of the Physical Society of Japan}, 81(1):011008, 2012.

\end{thebibliography}

\end{document}